\documentclass[preprint,prc,showpacs,preprintnumbers,amsmath,amssymb]{revtex4-1}
\usepackage{graphicx,latexsym,amssymb,amsmath,color,multirow,mathrsfs,booktabs, 
ifpdf}
\usepackage{dcolumn}
\usepackage{bm}
\usepackage{longtable}

\def\aC{$\alpha+^{12}$C\ }
\def\aap{($\alpha,\alpha'$)\ }
\def\ppp{($p,p'$)\ }
\def\aA{$\alpha$-nucleus\ }

\begin{document}
\title{Folding model analysis of the inelastic \aC scattering at medium
energies, and the isoscalar transition strengths of the cluster states 
of $^{12}$C}
\author{Do Cong Cuong$^1$}
\author{Dao T. Khoa$^1$\footnote{khoa@vinatom.gov.vn}}
\author{Yoshiko Kanada-En'yo$^2$}
\affiliation{$^1$Institute for Nuclear Science and Technology, VINATOM \\
179 Hoang Quoc Viet Rd., Hanoi, Vietnam.\\
$^2$ Department of Physics, Kyoto University, Kyoto 606-8502, Japan. }
\begin{abstract}
\begin{description}
\item[Background] 
The (spin and isospin zero) $\alpha$-particle is an efficient projectile 
for the excitation of the isoscalar, natural-parity states of $^{12}$C. 
Among those states that have pronounced $\alpha$-cluster structure, the 
Hoyle state ($0^+_2$ state at 7.65 MeV) has been observed in many \aap
experiments while the second $2^+$ state of $^{12}$C, predicted at 
$E_{\rm x}\approx 10$ MeV as an excitation of the Hoyle state, has not 
been observed until a recent high-precision experiment of the \aC scattering 
at $E_\alpha=386$ MeV. A plausible reason is a strong population 
of the narrow $3^-_1$ state at 9.64 MeV and broad 0$^+_3$ resonance at 
10.3 MeV that hinder the weak $2^+_2$ peak in the \aap spectrum.   
\item[Purpose] The accurate determination of the electric $E\lambda$
transition strengths of the isoscalar states of $^{12}$C, including a 
$E2$ component at $E_{\rm x}\approx 10$ MeV that can be assigned to the 
$2^+_2$ state, based on a detailed folding model + coupled channel analysis 
of the \aap data measured at $E_\alpha=240$ and 386 MeV.  
\item[Method] The complex optical potential and inelastic form factor 
given by the folding model for the \aC scattering are used to calculate 
the \aap cross sections for the known isoscalar states of $^{12}$C in 
an elaborate coupled channel approach. The strengths of the form factors 
for these states are then fine tuned against the \aap data to deduce the 
corresponding $E\lambda$ transition strengths.  
\item[Results] A significant $E2$ transition strength has been obtained for 
the $2^+_2$ state from the present analysis of the \aap data measured at 
$E_\alpha=240$ and 386 MeV. The $E\lambda$ transition strengths of the 
$0^+_2,\ 3^-_1,\ 0^+_3,$ and $1^-_1$ states were also carefully deduced, 
and some difference from the results of earlier analyses has been found 
and qualitatively understood.       
\item[Conclusion] Despite a strong hindrance by the $3^-_1$ and $0^+_3$ 
excitations, the presence of the $2^+_2$ state in the \aap spectra measured
at $E_\alpha=240$ and 386 MeV has been consistently confirmed by the present 
folding model + coupled channel analysis. 
\end{description}
\end{abstract}
\date{\today}

\maketitle

\section{Introduction}
\label{intro}
The excited states of $^{12}$C at energies near the $\alpha$-decay threshold 
have attracted a broad interest recently \cite{Fre07,Fun09} because of the 
dominant $\alpha$-cluster structure established in several cases, like that 
of the isoscalar 0$^+_2$ state at 7.65 MeV in $^{12}$C (known as the Hoyle 
state that plays a vital role in the carbon synthesis). Although a three 
$\alpha$-cluster structure of the Hoyle state has been shown more than 
three decades ago by the Resonating Group Method (RGM) calculations 
\cite{Ueg77,Kam81,Pic97}, an interesting $\alpha$-condensate scenario 
\cite{Fun09} for this state has been suggested recently \cite{Toh01,Fun03}, 
where three $\alpha$ clusters were shown to condense into the lowest $S$ 
state of their potential. Nevertheless, a more complicated structure of the 
Hoyle state is still being discussed \cite{Che07,Fre11}. Given a strongly 
nonspherical shape of $^{12}$C in the Hoyle state, an excited rotational 
band with the angular momentum $J^\pi = 2^+, 4^+,...$ built upon the Hoyle 
state has been suggested more than 50 years ago by Morinaga \cite{Mor56}. 
In the $\alpha$-condensate scenario, where the Hoyle state is the lowest $S$ 
state, it is also natural that the next level in the potential containing 
three $\alpha$-particles should be a $2^+$ state formed by promoting an 
$\alpha$-particle from the $S$ to $D$ level. The second 2$^+$ state 
of $^{12}$C has been predicted by several structure models 
\cite{Fun05,Kato,Enyo07,Epe12} at the excitation energy around 10 MeV, i.e.,
about 2 MeV above the $\alpha$-decay threshold, with a pronounced 
$^8$Be+$\alpha$ structure \cite{Fre07}. 

Because of such an interesting structure predicted for the 2$^+_2$ state 
of $^{12}$C, numerous experimental studies over the years have been aimed 
to detect it in the measured spectra of different reactions involving $^{12}$C 
(see, e.g., Refs.~\cite{Fre07a,Diget07,Fre09,Bri10,Zim11,Fre12}). The 
experimental observation of the $2^+_2$ state of $^{12}$C would be very 
important for a deeper understanding of the structure of the Hoyle state. 
In particular, the measured excitation energy would allow us to determine the 
moment of inertia and deformation of $^{12}$C being in the Hoyle state 
\cite{Mor56,Fre12,Fri71}. Although, some experimental evidence for a broad 
$2^+$ resonance was found in the mentioned experiments that might be assigned 
to the $2^+_2$ state of $^{12}$C, a clear identification of this state could 
be made just recently in the high-precision experiments on the inelastic 
$\alpha$ scattering \cite{Itoh11} and photodissociation of carbon 
\cite{Zim13,Zim13d}. The plausible explanations for the difficulty in 
identifying the $2^+_2$ state are:  

i) like all states above the $\alpha$-decay threshold, the $2^+_2$ state is
unstable against the disintegration of $^{12}$C$^*$ into three $\alpha$ 
particles and is, therefore, a short-lived resonance that is difficult 
to locate \cite{Fre11,Fri71}.
 
ii) there is always a strong population of the narrow $3^-_1$ state at 9.64 MeV 
and broad 0$^+_3$ resonance at 10.3 MeV that hinder the $2^+_2$ peak at 
about 10 MeV in the excitation spectrum of $^{12}$C \cite{Bri10}.   

We believe that the latter is the main reason why it was so difficult to
observe the $2^+_2$ state of $^{12}$C in the inelastic \aap or \ppp scattering. 
Our first attempt to investigate this puzzled situation based on a detailed 
folding model analysis of the inelastic \aC scattering data at 240 MeV 
\cite{John03} has been done in Ref.~\cite{Kho11}, where a weak $2^+_2$ peak 
at $E_x\approx 9\sim 10$ MeV has been shown to be strongly hindered by the 
$3^-_1$ peak at 9.64 MeV and 0$^+_3$ resonance at 10.3 MeV. Given a recent 
observation of the $2^+_2$ state of $^{12}$C in the high-precision \aap 
measurement at $E_\alpha=386$ MeV \cite{Itoh11} as well as its location and 
the $E2$ strength determined accurately from the photodissociation experiment 
\cite{Zim13,Zim13d}, we found it necessary to carry out again a consistent 
folding model analysis of the inelastic \aC scattering data measured at 
$E_\alpha=240$ and 386 MeV using the nuclear transition densities predicted 
by the antisymmetrized molecular dynamics (AMD) calculation \cite{Enyo07}. 
Our goal is not only to give a microscopic description of the \aap data at 
these two energies and try to deduce the $E2$ transition strength of the 
$2^+_2$ state of $^{12}$C from the experimental cross section at 
the excitation energy $E_x\approx 10$ MeV, but also to understand why the 
$2^+_2$ state could not be identified at this energy by the original 
multipole decomposition analysis of the 240 MeV \aap data \cite{John03}.

\section{Theoretical models}
\label{sec2}
\subsection{The double-folding model} 
\label{dfm}
The generalized double-folding model of Ref.~\cite{Kho00} was used to evaluate
the complex \aC optical potential (OP) and inelastic scattering form factor (FF)
from the Hartree-Fock type matrix elements of the (complex) effective nucleon-nucleon
($NN$) interaction between the projectile nucleon $i$ and target nucleon $j$
\begin{equation}
 U_{A\to A^*}=\sum_{i\in \alpha;j\in A,j'\in A^*}[\langle ij'|v_{\rm D}|ij\rangle
 + \langle ij'|v_{\rm EX}|ji\rangle],
\label{fd1}
\end{equation}
where $A$ and $A^*$ denote the target in the entrance- and exit channel
of the \aap scattering, respectively. Thus, Eq.~(\ref{fd1}) gives the (diagonal)
OP if $A^*=A$ and (nondiagonal) inelastic scattering FF if otherwise.
The (local) direct term is evaluated by the standard double-folding
integration 
\begin{eqnarray}
 U_{\rm D}(E,\bm{R})=\int\rho_\alpha(\bm{r}_\alpha)
 \rho_A(\bm{r}_A)v_{\rm D}(E,\rho,s)d^3r_\alpha d^3r_A,\nonumber \\
 \bm{s}=\bm{r}_A-\bm{r}_\alpha+\bm{R}.
\label{fd2}
\end{eqnarray}
The antisymmetrization gives rise to the exchange term in Eq.~(\ref{fd1}) which
is, in general, nonlocal. An accurate local equivalent exchange potential can be 
obtained \cite{Kho00,Kho07} using the local WKB approximation \cite{Sat83} for the 
change in relative motion induced by the exchange of spatial coordinates of each 
interacting nucleon pair
\begin{eqnarray}
U_{\rm EX}(E,\bm{R}) =\int \rho_\alpha
(\bm{r}_\alpha,\bm{r}_\alpha+\bm{s})\rho_A(\bm{r}_A,\bm{r}_A
-\bm{s})v_{\rm EX}(E,\rho,s) \nonumber \\
\times\exp\left({i\bm{K}(R)\bm{s}}\over{M}\right)d^3r_\alpha d^3r_A.
\label{fd3}
\end{eqnarray}
Here $\bm{K}(R)$ is the local momentum of relative motion determined as
\begin{equation}
 K^2(R)={{2\mu}\over{\hbar}^2}[E_{\rm c.m.}-{\rm Re}~U_0(E,R)-V_C(R)],
\label{fd4}
\end{equation}
where $\mu$ is the reduced mass, $M=4A/(4+A)$, $E_{\rm c.m.}$ is the scattering
energy in the center-of-mass (c.m.) frame, $U_0(E,R)$ and $V_C(R)$
are the nuclear and Coulomb parts of the \emph{real} OP, respectively. The
calculation of $U_{\rm EX}$ is done iteratively based on a density-matrix
expansion method \cite{Kho00,Kho01}. We have used here a realistic local 
approximation for the \emph{transition} density matrix suggested 
by Love \cite{Lo78}. The recoil correction to the exchange term (\ref{fd3}) 
suggested by Carstoiu and Lassaut \cite{Car96} has been taken into account.

Among different choices of the effective $NN$ interaction, a density dependent
version of the M3Y-Paris interaction (dubbed as CDM3Y6 interaction \cite{Kho97}) 
has been used quite successfully in the folding model analyses of the elastic
and inelastic \aA scattering \cite{Kho07}. The density dependent parameters 
of the CDM3Y6 interaction were carefully adjusted in the Hartree-Fock 
scheme to reproduce the saturation properties of nuclear matter \cite{Kho97}. 
To avoid a phenomenological choice of the imaginary parts of the OP and 
inelastic FF, we have supplemented the M3Y-Paris interaction with a realistic 
\emph{imaginary} density dependence \cite{Kho10} for the folding calculation of
the imaginary parts of the OP and inelastic FF. The parameters of the imaginary 
density dependence have been deduced at each energy based on the Brueckner 
Hartree-Fock results for the nucleon OP in nuclear matter by Jeukenne, Lejeune 
and Mahaux (the well-known JLM potential \cite{Je77}). The explicit density 
dependent parameters of the (complex) CDM3Y6 interaction for the \aA scattering 
at the energies $E_\alpha=240$ and 386 MeV are given in Ref.~\cite{Kho10}.

The key quantity in our folding model analysis of the inelastic \aA scattering 
is the inelastic FF that contains all the structure information of the
nuclear state under study. Given an accurate choice of the effective $NN$ 
interaction, the present double-folding approach can be applied successfully to 
study the inelastic \aC scattering only if the realistic nuclear densities were 
used in the folding calculation (\ref{fd2})-(\ref{fd3}). In our earlier 
studies, the nuclear densities given by the RGM wave functions \cite{Kam81} 
have been used in the folding model analysis to probe the $E0$ transition 
strength of the Hoyle state \cite{Kho08}, and other isoscalar excitations 
of $^{12}$C, like 2$^+_1$ (4.44 MeV), 3$^-_1$ (9.64 MeV), 0$^+_3$ (10.3 MeV) 
and 1$^-_1$ (10.84 MeV) states \cite{Kho08k}. Like Ref.~\cite{Kho11}, the 
nuclear densities given by the AMD approach \cite{Enyo07} have been used in 
the present folding model analysis of the inelastic \aC scattering at 
$E_\alpha=240$ and 386 MeV. 

\subsection{The AMD nuclear transition densities in the present folding
 model analysis}
\label{amd}
The AMD approach was proven to give quite realistic description of the 
structure of the low-lying states in light nuclei, where both the cluster and 
shell-model like states are consistently reproduced \cite{Enyo07}. In the 
present work, the isoscalar states of $^{12}$C are generated by the AMD 
approach using the method of variation after the spin-parity projection. 
The main structure properties of these states are summarized in Table~\ref{t1}.

\begin{table}\centering\vspace*{-0.5cm}
\caption{Excitation energies and $E\lambda$ transition strengths of the IS
states of $^{12}$C under study. The calculated values are the AMD 
results \cite{Enyo07}, and the best-fit transition rates are given by 
the present folding model + CC analysis. $M(E\lambda)$ is given in 
$e$~fm$^{\lambda+2}$ for the 0$^+$ and 1$^-$ states.}
\vspace*{0.5cm}
\begin{tabular}{|c|c|c|c|c|c|c|c|c|} \hline
$J^{\pi}$ & $\langle r^2 \rangle^{1/2}_{\rm calc}$ & $E_{\rm calc}$ & $
 E_{\rm exp}$ & Transition & calc. & best-fit & exp. & Ref. \\
 & (fm) & (MeV) & (MeV) &  & ($e^2$fm$^{2\lambda}$) & ($e^2$fm$^{2\lambda}$) & 
 (e$^2$fm$^{2\lambda}$) & \\ \hline
 2$^+_1$ & 2.668 & 4.5 & 4.44 &$B(E2;2^+_1\rightarrow 0^+_1)$ & 8.4 & 
 $8.4\pm 1.5$ &  $7.4\pm 0.2$ & \cite{Itoh11} \\
  &  &  &  & & & & $7.7\pm 1.0$ & \cite{John03} \\
  &  &  &  & & & & $8.0\pm 0.8$ &  \cite{Ram01} \\
  &  &  &  & $B(E2;2^+_1\rightarrow 4^+_1)$ & 28.5  & &  & \\ \hline
 0$^+_2$ & 3.277 & 8.1 & 7.65 & $M(E0;0^+_2\rightarrow 0^+_1)$ & 6.6 & 
 $4.5\pm 0.5$ & $3.7\pm 0.2$ & \cite{John03} \\ 
  &  &  &  & & & & $5.4\pm 0.2$ & \cite{Stre70} \\ 
  &  &  &  &$B(E2;0^+_2\rightarrow 2^+_1)$ & 25.5 & & $13.0\pm 2.0$  &
	\cite{Endt79} \\
	&  &  &  & $B(E3;0^+_2\rightarrow 3^-_1)$ & 3122 & & & \\  
  &  &  &  & $M(E0;0^+_2\rightarrow 0^+_3)$ & 16.7 & & & \\  
  &  &  &  & $M(E1;0^+_2\rightarrow 1^-_1)$ & 12.5  & & & \\  \hline
3$^-_1$ & 3.139 & 10.8 & 9.64 & $B(E3;3^-_1\rightarrow 0^+_1)$ & 74.4 & 
$59.5\pm 3.2$ & $35.9\pm 1.4$ & \cite{Itoh11} \\
 &  &  &  & & & & $34.3\pm 5.7$ & \cite{John03} \\
 &  &  &  & & & &  $87.1\pm 1.3$ & \cite{Kib02} \\
 &  &  &  &$B(E3;3^-_1\rightarrow 2^+_2)$ & 136.7 &  &  &  \\
 &  &  &  &$M(E1;3^-_1\rightarrow 2^+_2)$ &3.71 &  &  &  \\ \hline
0$^+_3$ & 3.985 & 10.7 & 10.3 & $M(E0;0^+_3\rightarrow 0^+_1)$ & 2.3 & 
$2.9\pm 0.3$ & $3.0\pm 0.2$ & \cite{John03} \\
 &  &  &  & $B(E2;0^+_3\rightarrow 2^+_2)$ & 1553 & & & \\  \hline
2$^+_2$ & 3.993 & 10.6 & 9.84 & $B(E2;2^+_2\rightarrow 0^+_1)$ & 
0.4 & $0.6\pm 0.1$ & $0.37\pm 0.02$ & \cite{Itoh11} \\
 &  &  & 10.13 & & &  & $1.57\pm 0.14$ & \cite{Zim13,Zim13d} \\
 &  &  &  & $B(E2;2^+_2\rightarrow 0^+_2)$ &  102 &  & & \\
 &  &  &  & $B(E2;2^+_2\rightarrow 4^+_1)$ &  13.5 & & & \\
 &  &  &  & $B(E2;2^+_2\rightarrow 4^+_2)$ &  1071 &  & & \\  \hline
1$^-_1$ & 3.424 & 12.6 & 10.84 & $M(E1;1^-_1\rightarrow 0^+_1)$ & 1.58 & 
$0.34\pm 0.04$ & $0.31\pm 0.04$ & \cite{John03} \\
 &  &  &  & $M(E1;1^-_1\rightarrow 2^+_2)$ & 3.73 & & &  \\
 &  &  &  & $B(E3;1^-_1\rightarrow 2^+_2)$ & 1679   & & & \\ \hline
\end{tabular}\label{t1} \\
\end{table}

While the AMD prediction for the shell-model like $2^+_1$ state is quite 
satisfactory in both the excitation energy and $E2$ transition strength, the 
predicted excitation energies for higher lying states are larger than the 
experimental values. However, such a difference in the excitation energies 
leads only to a very small change in the kinetic energy of emitted 
$\alpha$-particle and does not affect significantly the inelastic \aC 
scattering cross sections calculated in the Distorted Wave Born 
Approximation (DWBA) or coupled-channel (CC) formalism. On the other hand, 
the strength and shape of the nuclear transition density used to evaluate 
the inelastic FF are the most vital inputs that affect directly the 
calculated \aap cross section. The details of the AMD approach 
to the excited states of $^{12}$C are given in Ref.~\cite{Enyo07}. In the 
present work, the AMD nuclear transition densities enter the folding calculation 
in the same convention as that used in Refs.~\cite{Kho00,Kho10, Kho11} so that 
the isoscalar (IS) transition strength for a $2^\lambda$-pole nuclear transition
$|J_i\rangle\to |J_f\rangle$ is described by the reduced nuclear transition rate
$B({\rm IS}\lambda;J_i\to J_f)=|M({\rm IS}\lambda;J_i\to J_f)|^2$, where the
$2^\lambda$-pole transition moment is determined from the corresponding nuclear 
transition density as
\begin{eqnarray}
M({\rm IS}\lambda;J_i\to J_f)&=&\int dr\ r^{\lambda+2}
 \rho_{J_f,J_i}^{(\lambda)}(r) \ \ \  {\rm if}\ \ \  \lambda\geqslant 2, 
 \label{dens} \\
M({\rm IS}0;J_i\to J_f)&=&\sqrt{4\pi}\int dr\ r^{4} \rho_{J_f,J_i}^{(\lambda=0)}(r),
 \label{dens0} \\
M({\rm IS}1;J_i\to J_f)& =& \int dr\left(r^{3}-
 {5\over 3}\langle r^2\rangle r\right)r^2 \rho_{J_f,J_i}^{(\lambda=1)}(r).  
 \label{dens1}
\end{eqnarray}
Note that the IS dipole transition moment is evaluated based on higher-order 
corrections to the dipole operator, with spurious c.m. oscillation subtracted 
\cite{giai}. The reduced electric transition rate is evaluated as
$B(E\lambda;J_i\to J_f)=|M(E\lambda;J_i\to J_f)|^2$, where $M(E\lambda)$ is 
determined in the same way as $M({\rm IS}\lambda)$ but using the proton part 
of the nuclear transition density only. We will discuss hereafter the 
transition strength in terms of $B(E\lambda)$, because this is the quantity 
that can be compared with the experimental data. The excitation energies and 
$E\lambda$ transition strengths of the excited states of $^{12}$C are given 
in Table~\ref{t1}. One can see that the AMD results for the excitation 
energies and $E\lambda$ transition strengths between the ground state $0^+_1$ 
and the excited $2^+_1$, $0^+_2$ and $3^-_1$ states agree 
reasonably with the experimental values.

In difference from the shell-model like $2^+_1$ state, the $2^+_2$ state has a 
well defined cluster structure (see Fig.~5 of Ref.~\cite{Enyo07}), with a more 
extended mass distribution that leads to the matter radius 
$R_{\rm m}=\langle r^2 \rangle^{1/2}\approx 3.99$ fm which is even larger than 
that predicted for the Hoyle state. The more interesting are the predicted 
$E2$ transitions from the Hoyle state to the $2^+_2$ state and from the $2^+_2$ 
state to the $4^+_2$ state, $B(E2;0^+_2\rightarrow 2^+_2)\approx 511\ e^2$fm$^4$ 
and $B(E2;2^+_2\rightarrow 4^+_2)\approx 1071\ e^2$fm$^4$ that are much stronger 
than the $E2$ transitions between the members of the ground-state band,
$B(E2;0^+_1\rightarrow 2^+_1)\approx 42.5\  e^2$fm$^4$
and $B(E2;2^+_1\rightarrow 4^+_1)\approx 28.5\  e^2$fm$^4$. Thus, the $E2$
transition rates predicted by the AMD strongly suggest that the $2^+_2$ and 
$4^+_2$ states should be the members of a rotational band built upon the 
Hoyle state. We note that the $B(E2;0^+_2\rightarrow 2^+_2)$ value predicted 
by the RGM \cite{Kam81} or fermion molecular dynamics calculations \cite{Neff} 
is even larger than that given by the AMD. The direct excitation of the $2^+_2$ 
state from the ground state has been predicted to be very weak, with 
$B(E2;0^+_1\rightarrow 2^+_2)\approx 2\  e^2$fm$^4$ that is significantly 
smaller than the latest experimental value deduced from the photodissociation 
experiment $^{12}$C($\gamma,\alpha)^8$Be \cite{Zim13,Zim13d}. We note here that 
the original analysis of the photodissociation data  \cite{Zim13} resulted on 
$B(E2\uparrow)\approx 3.65\ e^2$fm$^4$ and the total width $\Gamma\approx 0.8$ 
MeV for the $2^+_2$ state. However, with some more data points taken, the 
revised analysis of the $^{12}$C($\gamma,\alpha)^8$Be data has given 
$B(E2\uparrow)\approx 7.85\ e^2$fm$^4$ and $\Gamma\approx 2.1$ MeV \cite{Zim13d}. 
 
\begin{figure}
 \begin{center}
\vspace*{-2cm}\hspace*{0cm}
\includegraphics[angle=0,scale=0.60]{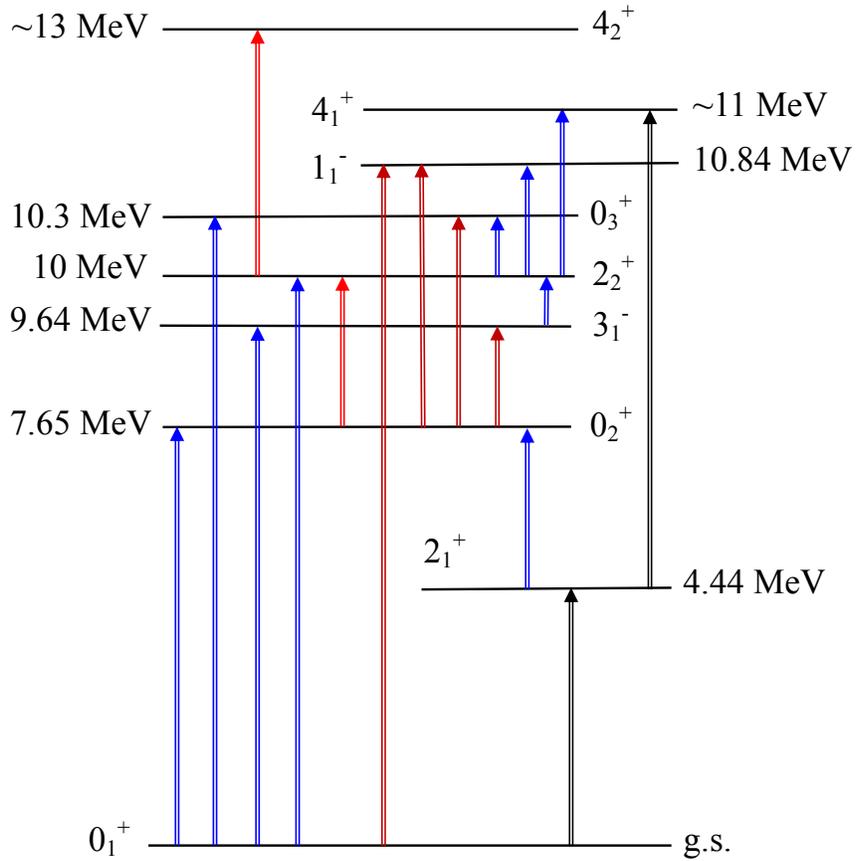} \vspace{-2cm}
\caption{(Color online) Coupling scheme used in the present coupled channel 
calculation of the elastic and inelastic \aC scattering.}\label{cc}
 \end{center}
\end{figure}

Nevertheless, the newly found $B(E2\uparrow)$ value for the $2^+_2$ state is 
still at least 5 times weaker than the $B(E2\uparrow)$ value established for 
the $2^+_1$ state \cite{Ram01}. Therefore, the $2^+_2$ state must be a very weak 
(\emph{direct}) excitation of $^{12}$C and it is, therefore, difficult to 
observe this state in the inelastic hadron scattering. Although a very strong 
$E2$ transition has been predicted for the excitation of the $2^+_2$ state 
from the Hoyle state, such a two-step excitation of $^{12}$C via the Hoyle 
state seems to be suppressed at the medium- and high incident $\alpha$ 
energies as well as by the disintegration of the excited $^{12}$C$^*$ into 
three $\alpha$ particles. This could be also another reason for the 
scarcity of the experimental observation of the $2^+_2$ state. 
Beside a strong $E2$ transition between the Hoyle state and $2^+_2$ state
discussed above, AMD also predicted very strong $E\lambda$ transitions 
from the Hoyle state to the $3^-_1$, $0^+_3$ and $1^-_1$ states 
that are about order of magnitude larger than $E\lambda$ transitions 
between these states and the ground state (see Table~\ref{t1}). To
have an accurate coupled channel scenario for the \aap scattering under
study, all $E\lambda$ transitions shown in Table~\ref{t1} have been 
included into the present CC calculation, and the coupling scheme
in Fig.~\ref{cc} is, therefore, more comprehensive than that used
earlier in Refs.~\cite{Kho11,Kho08}. 

\begin{figure}
 \begin{center}
\vspace*{-1cm}\hspace*{0cm}
\includegraphics[angle=0,scale=0.6]{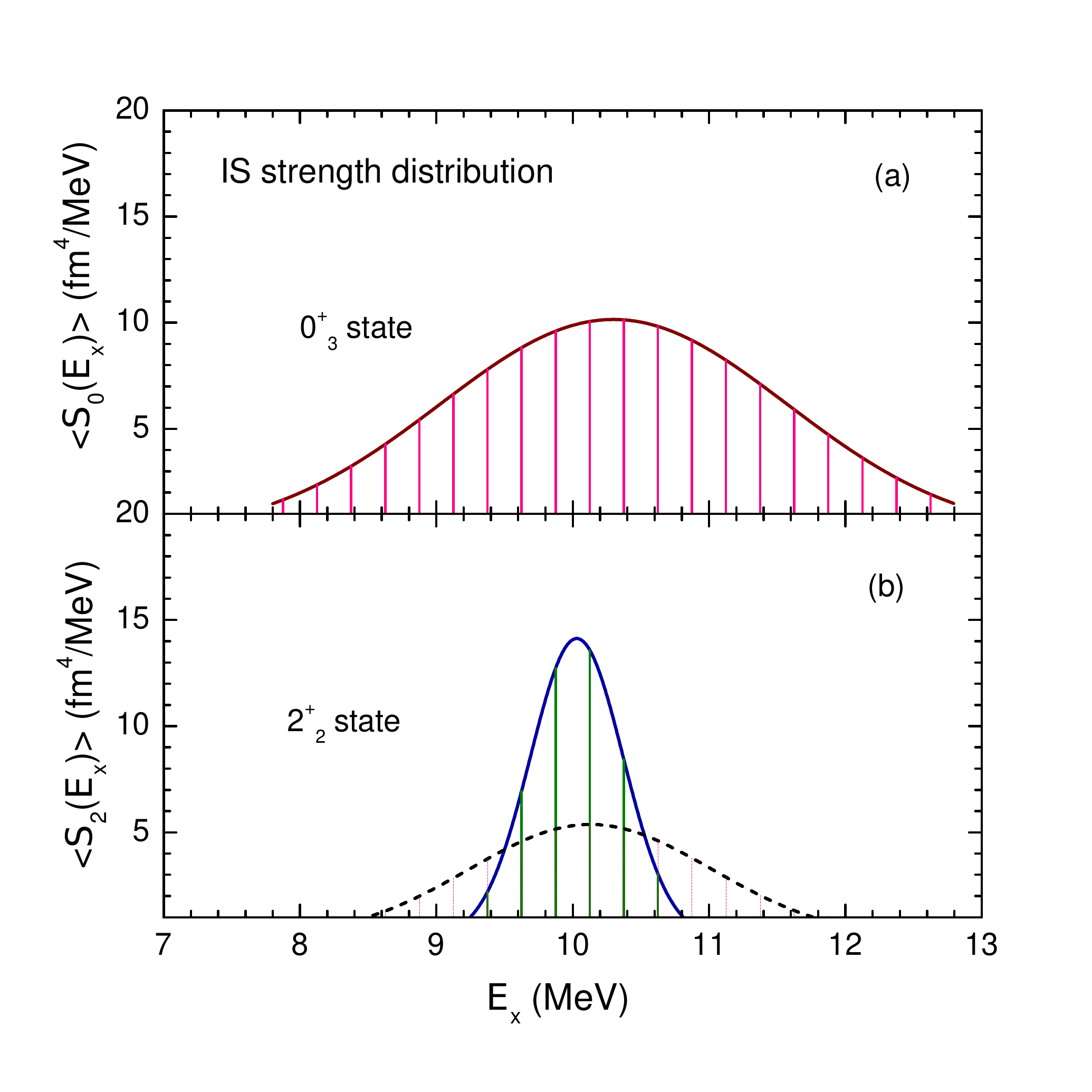} \vspace{-1cm}
\caption{(Color online) Distribution of the isoscalar transition strength 
(\ref{av1}) of the $0^+_3$ state (a) and $2^+_2$ state (b) over the 250 keV-wide 
energy bins around $E_x\sim 10$ MeV used in the present folding model + CC 
analysis of the \aap data measured at $E_\alpha= 386$ MeV. The $E2$ strength
distributions of the $2^+_2$ state based on the total width $\Gamma=0.8$ and 
2.1 MeV are shown as solid and dash lines, respectively.}\label{str}
 \end{center}
\end{figure}

In the inelastic \aC scattering experiments at $E_\alpha=240$ \cite{John03}
and 386 MeV \cite{Itoh11}, the \aap cross sections have been measured 
accurately in small energy bins of 475 and 250 keV width, respectively, 
over a wide range of scattering angles and excitation energies. These data 
have been subjected to the multipole decomposition analysis (MDA) to 
disentangle contribution of different $E\lambda$ multipolarities to the 
excitation of $^{12}$C in each energy bin.  
The MDA technique is the same as that used earlier to deduce the electric 
transition strengths of the isoscalar giant resonances from \aap spectrum. 
As a result, the \aap cross sections measured at $E_\alpha=386$ MeV for the 
energy bins centered around $E_x\sim 10$ MeV were shown to contain the 
contribution from both the $0^+_3$ and $2^+_2$ states of $^{12}$C \cite{Itoh11}. 
Although the $3^-_1$ state at 9.64 MeV has a very narrow width of about 34 keV, 
the observed $3^-$ strength is distributed over a much wider energy range 
of the \aap spectrum that is likely associated with the energy resolution 
of the measurement (the energy resolution of the 386 MeV measurement is 
around 200 keV \cite{Itoh11}). 
Given our best-fit $B(E3)$ transition strength comparable to that 
deduced from the $(e,e')$ experiment and a good description of the 
measured $3^-_1$ angular distribution by the inelastic FF obtained with  
the AMD transition density (see Table~\ref{t1} and Fig.~\ref{inel3}), 
we have scaled the $3^-_1$ transition density to reproduce the $E3$ 
strength found in each energy bin by the MDA of the \aap data 
\cite{Itoh11,John03} and used it to calculate the $3^-$ inelastic FF 
for that energy bin. Situation with the $0^+_3$ and $2^+_2$ states is more 
uncertain, especially, no $E2$ strength was found by the MDA of the 
240 MeV data \cite{John03} in the energy bins centered around 10 MeV.
Therefore, we chose not to scale the AMD transition density to the 
$E0$ and $E2$ strengths given by the MDA of the \aap data and adopted the 
strength \emph{averaging} procedure \cite{Fes92}, used in our earlier folding 
model analysis of the \aap scattering on the lead target \cite{Kho10} at 
the same incident $\alpha$ energies, to predict the strength distribution 
of the $0^+_3$ and $2^+_2$ states over the considered energy bins. Namely, 
the IS transition strength $S_{\rm AMD}\equiv B({\rm IS}\lambda;J_i\to J_f)$, 
given by the AMD transition density scaled to give the best CC fit of the 
\aap data for the $0^+_3$ or $2^+_2$ states, has been spread over the 
excitation energy as
\begin{equation}
 \langle S(E)\rangle = S_{\rm AMD}f(E-E_x), \label{av1}
\end{equation}
where the adopted experimental excitation energies $E_x\approx 10.3$ and 10 MeV 
\cite{Itoh11,John03,Zim13} have been used for the $0^+_3$ and $2^+_2$ states, 
respectively, and   
\begin{equation}
 S_{\rm AMD} =\int_E \langle S(E')\rangle dE'. \label{av2}
\end{equation}
The averaging function $f(E-E_{x})$ is a Gaussian distribution 
 \begin{equation}
f(E-E_{x})=\frac{1}{\sigma\sqrt{2\pi}}
\exp\left(-\frac{1}{2}\left(\frac{E-E_{x}}{\sigma}\right)^2\right), 
\label{av3}\end{equation}
where $\sigma$ is the standard deviation associated with the full width $\Gamma$ 
at half maximum as $\Gamma=2.355\sigma$. The obtained IS transition strength 
distribution of the $0^+_3$ and $2^+_2$ states is shown in Fig.~\ref{str}, 
where the adopted experimental width $\Gamma\approx 3$ MeV \cite{Itoh11,John03} 
has been used for the $0^+_3$ state. The total width of the $2^+_2$ state has
been suggested as $\Gamma\approx 0.7\sim 0.8$ MeV by most of the experimental 
studies \cite{Fre07a,Diget07,Fre09,Bri10,Itoh11,Zim13}. However, the revised
analysis of the $^{12}$C($\gamma,\alpha)^8$Be data has ``unambiguously" 
determined $\Gamma\approx 2.1\pm 0.3$ MeV for the $2^+_2$ state \cite{Zim13d}.
To deal with such a situation, we have used two different total widths 
$\Gamma=0.8$ and 2.1 MeV as the inputs for the averaging procedure (\ref{av1}) 
of the $E2$ strength of the $2^+_2$ state (see lower panel of Fig.~\ref{str}).

To obtain the inelastic scattering FF for the contribution of the considered state
in each energy bin, we have used for the input of the double-folding calculation 
(\ref{fd2})-(\ref{fd3}) the bin transition density of this state determined as
 \begin{equation}
 \rho^{(\lambda)}_{\rm bin}(r)=\sqrt{\frac{S_{\rm bin}}{S_{\rm AMD}}}
 \rho^{(\lambda)}_{\rm AMD}(r), \label{av4}
\end{equation}
where $\rho^{(\lambda)}_{\rm AMD}(r)$ is the AMD transition density of the 
$0^+_3$ or $2^+_2$ state, scaled to give the best CC fit of the \aap data, 
and $S_{\rm bin}$ is the IS transition strength in the energy bin
\begin{equation}
 S_{\rm bin} =\int_{E_{\rm bin}-\Delta E}^{E_{\rm bin}+\Delta E} 
 \langle S(E')\rangle dE'. \label{av5}
\end{equation}
Here $E_{\rm bin}$ is the center of the energy bin and $\Delta E$ is its
half width.

\section{Results and discussion}
\label{sec3}
Given the AMD transition densities calculated for different transitions 
between the IS states of $^{12}$C shown in Table~\ref{t1}, the corresponding 
inelastic folded FF can be used in both the DWBA or CC analysis of the \aap 
data. For this kind of analysis, an accurate determination of the distorted 
waves in the entrance and exit channels by the appropriately chosen 
optical potential is very crucial. In the present work, the complex OP 
for the entrance channel is given by the double-folding calculation 
(\ref{fd2})-(\ref{fd3}) using the ground-state (g.s.) density of $^{12}$C
and complex CDM3Y6 density dependent interaction. 
Because the exit channel of the inelastic \aC scattering contains 
$^{12}$C$^*$ being in an excited (cluster) state that is generally more 
dilute, with the predicted $\langle r^2 \rangle^{1/2}$ radius significantly 
larger than $\langle r^2 \rangle^{1/2}_{\rm g.s.}\approx 2.33$ fm (see 
Table~\ref{t1}), the OP of each exit channel has been computed separately 
at the energy $E_\alpha-Q$, using the diagonal density of $^{12}$C$^*$ 
given by the AMD. It can be seen from the discussion below that such an 
elaborate treatment of the exit OP lead to a better agreement of the 
calculated \aap cross sections with the data at large angles, and helped 
to deduce accurately the $E\lambda$ transition rates for the considered 
excited states (the best-fit values given in Table~\ref{t1}).    

\begin{figure}
 \begin{center}\vspace*{-1cm}\hspace*{-1cm}
\includegraphics[angle=0,scale=0.8]{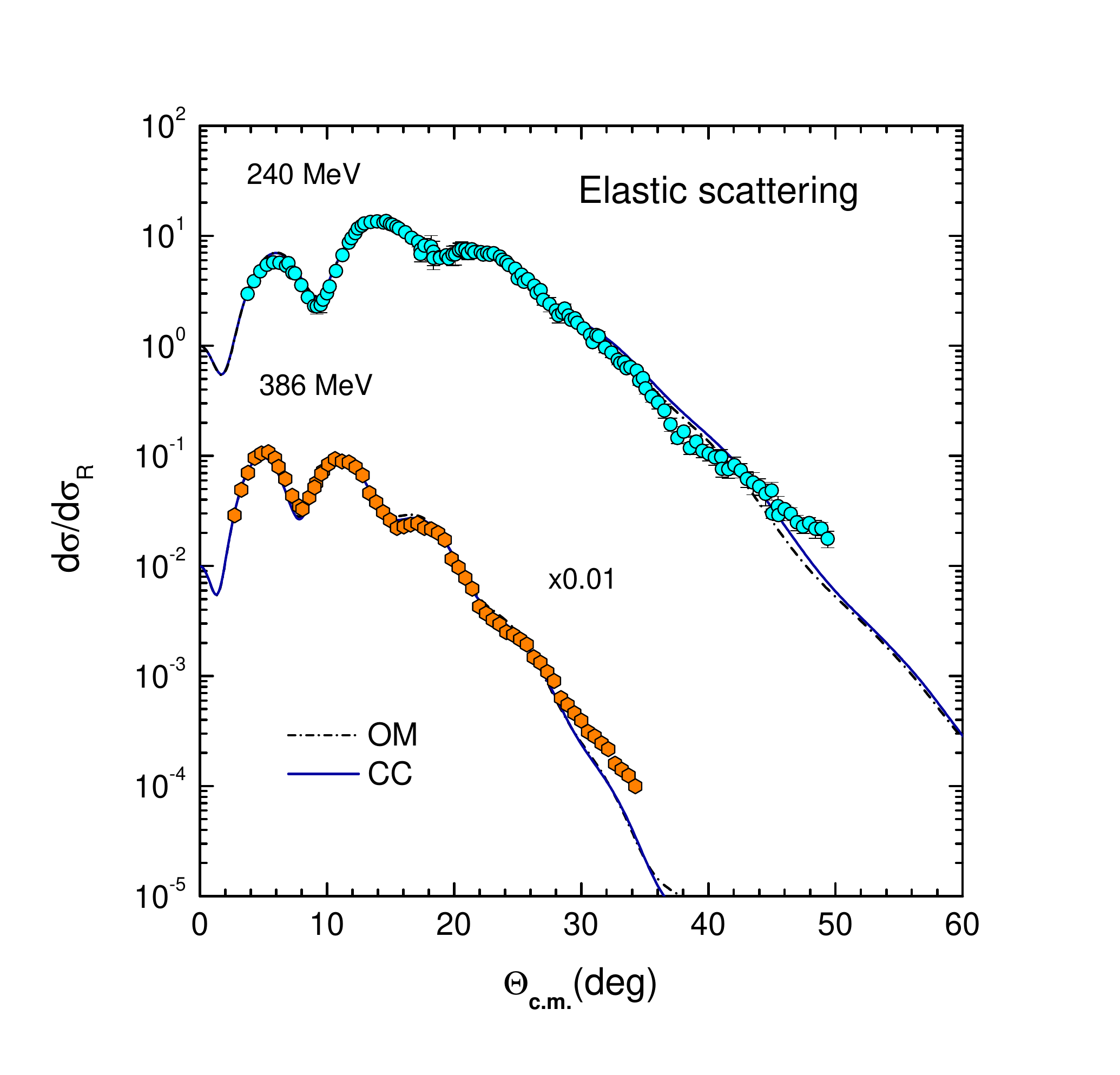}\vspace{-1cm}
\caption{(Color online) OM and CC descriptions of the elastic 
 \aC scattering data measured at $E_\alpha= 240$ MeV \cite{John03} 
 and 386 MeV \cite{Itoh11}.}\label{elastic}
 \end{center}
\end{figure}
All the optical model (OM), DWBA and CC calculations have been performed
using the code ECIS97 written by Raynal \cite{Raynal}. To account for the 
higher-order (dynamic polarization) contributions to the folded OP 
\cite{Kho00} and to fine tune the complex strength of the CDM3Y6 
interaction for each energy, the real and imaginary folded OP's were 
scaled by the coefficients $N_{\rm R}$ and $N_{\rm I}$, respectively, 
which were adjusted to the best OM fit of the elastic scattering data 
(see Fig.~\ref{elastic}). As a result, the best-fit $N_{\rm R}\approx 1.05,\ 
N_{\rm I}\approx 1.27$ and $N_{\rm R}\approx 1.24,\ N_{\rm I}\approx 1.38$ 
were obtained for $E_\alpha= 240$ and 386 MeV, respectively.
We note that the imaginary strength of the CDM3Y6 interaction was tuned to
the JLM results for nuclear matter and gives, therefore, only the ``volume"
absorption. To effectively account for the surface absorption caused by
inelastic scattering and transfer reactions etc., an enhanced $N_{\rm I}$ 
coefficient is naturally expected. The OM calculation using the complex
folded OP gives the total reaction cross section $\sigma_{\rm R}\approx
626$ and 624 mb for $E_\alpha= 240$ and 386 MeV, respectively, which are 
close to the $\sigma_{\rm R}$ value around 240 mb given by the empirical 
global OP for the elastic \aA scattering \cite{Anders00}.
These same $N_{\rm R(I)}$ factors were used to scale the real and imaginary 
inelastic folded FF for the DWBA calculation, a standard method widely 
adopted in the folding model + DWBA analysis of inelastic \aA scattering 
\cite{John03,Kho00,Sat97}. In the CC calculation, $N_{\rm R}$ and $N_{\rm I}$ 
must be re-adjusted again to account only for higher-order effects caused 
by the nonelastic channels not included into the CC scheme shown in 
Fig.~\ref{cc}. We have then obtained  $N_{\rm R}\approx 1.08,\ 
N_{\rm I}\approx 1.18$ and $N_{\rm R}\approx 1.26,\ N_{\rm I}\approx 1.35$ 
from the CC analysis of the 240 and 386 MeV data, respectively. These 
$N_{\rm R(I)}$ factors were used to scale the OP and all the (complex) 
inelastic folded FF's used in the present CC analysis of the \aap data. 
\begin{figure}
 \begin{center}\vspace*{-1cm}\hspace*{-1cm}
\includegraphics[angle=0,scale=0.6]{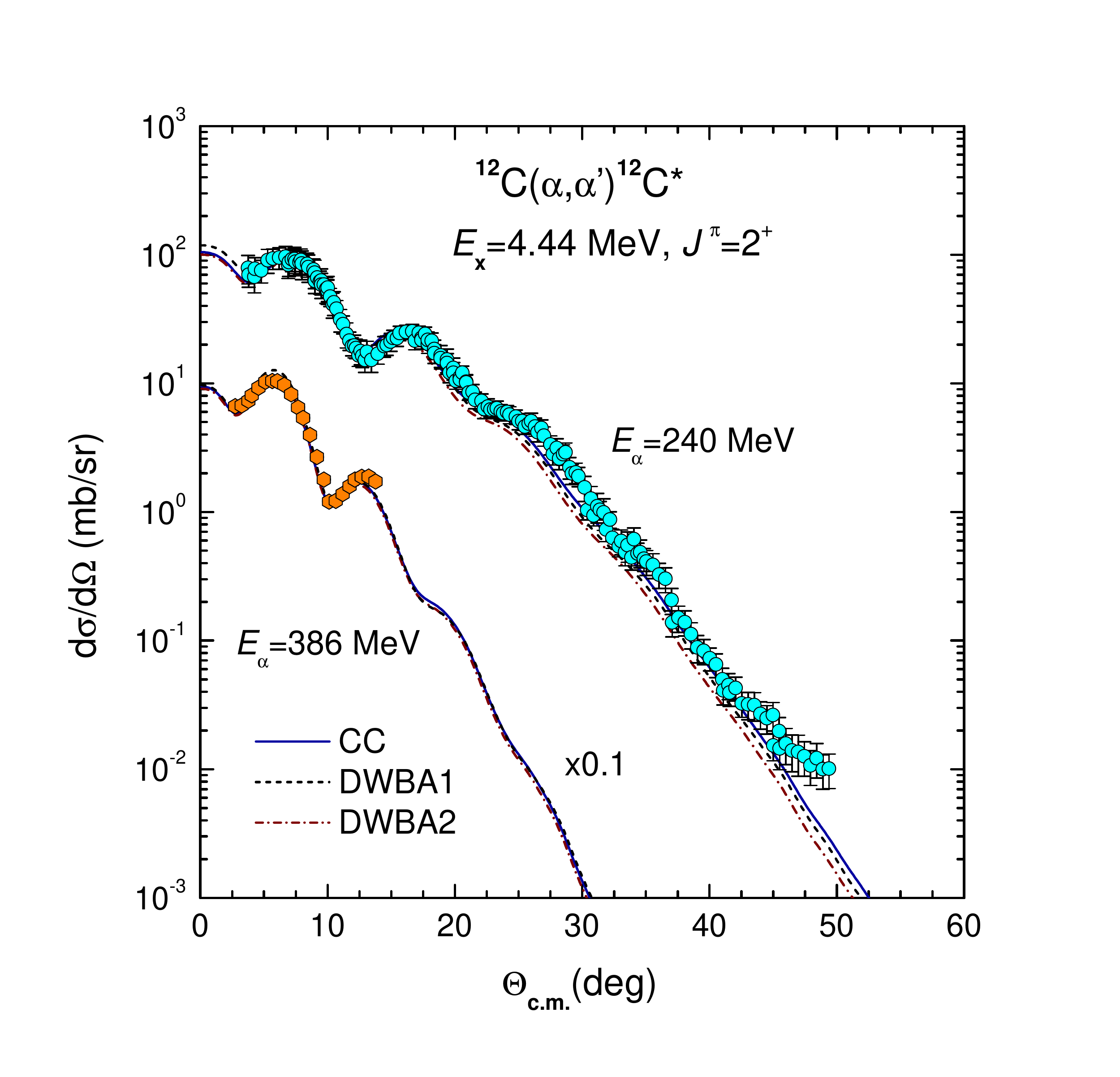}\vspace{-1cm}
\caption{(Color online) DWBA and CC descriptions of the inelastic \aC 
scattering data for the $2^+_1$ state, measured at $E_\alpha= 240$ MeV 
\cite{John03} and 386 MeV \cite{Itoh11}. The DWBA1 results were obtained 
using the same OP for both the entrance and exit channels, and the DWBA2 and
CC results were obtained with the OP of the exit channel computed separately 
at the energy $E_\alpha-Q$, using the AMD diagonal density of $^{12}$C$^*$.}
\label{inel2}
 \end{center}
\end{figure}
From the OM and CC results shown in Fig.~\ref{elastic} one can see
that the renormalized (complex) folded OP describes the elastic data 
accurately up to the large angles, and thus providing the realistic 
distorted waves for the DWBA and CC calculations of inelastic scattering.  
\begin{figure}
 \begin{center}\vspace*{-1cm}\hspace*{-1cm}
\includegraphics[angle=0,scale=0.6]{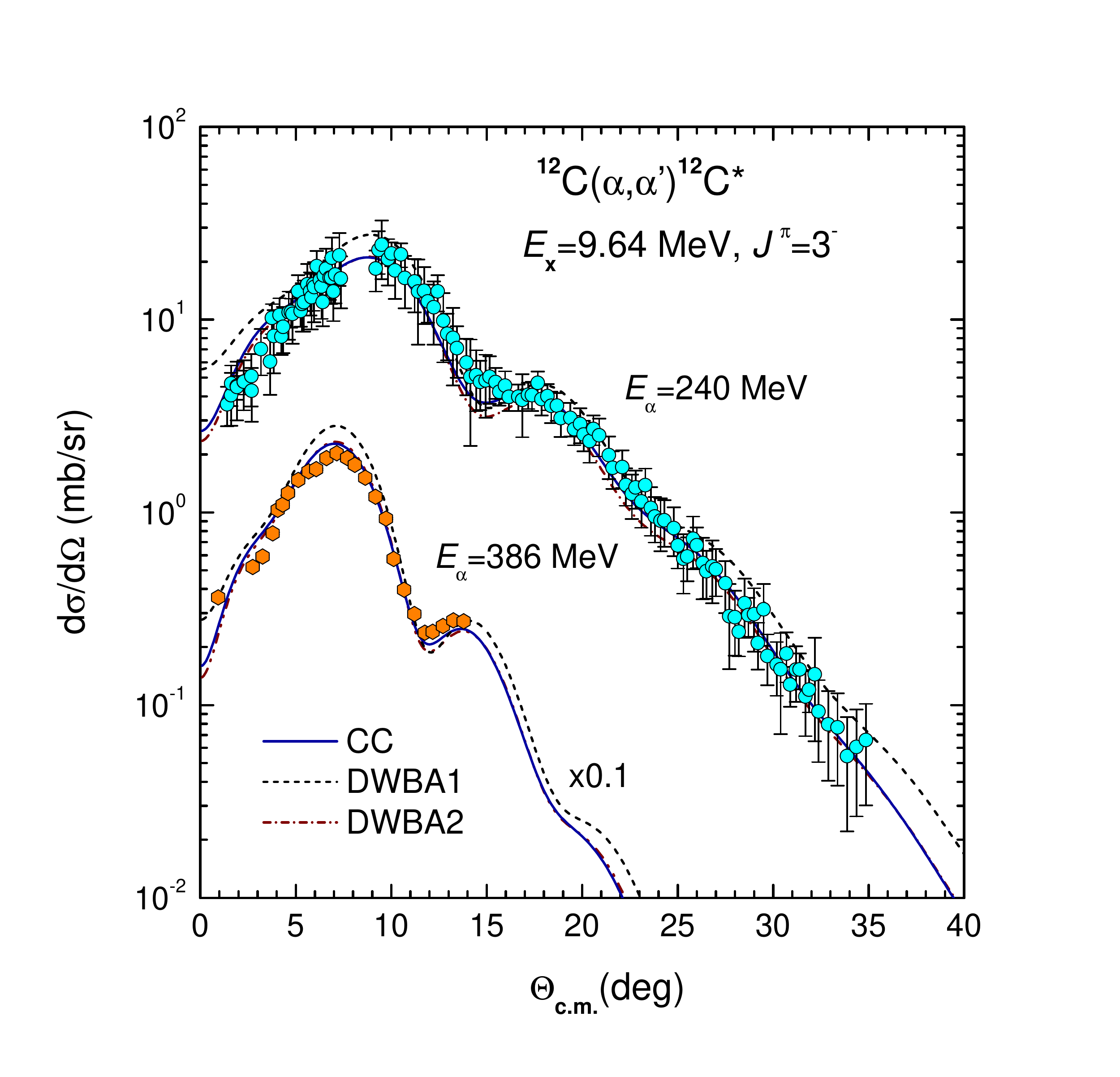}\vspace{-1cm}
\caption{(Color online) The same as Fig.~\ref{inel2} but for the $3^-_1$ state.}
\label{inel3}
 \end{center}
\end{figure}

The DWBA and CC results for the inelastic \aap scattering to the $2^+_1$ and 
$3^-_1$ states are compared with the experimental data in Figs.~\ref{inel2} 
and  \ref{inel3}, respectively. One can see that all the calculated \aap cross 
sections for the shell-model like $2^+_1$ state agree reasonably with the 
data, with the CC calculation giving a slightly better fit to the 240 MeV data 
at large angles. The $B(E2)$ transition strength predicted by the AMD agrees 
well with the experimental values deduced from the \aap and $(e,e')$ data 
\cite{Itoh11,John03,Ram01}, and the AMD nuclear transition density of the
$2^+_1$ state is well suitable for the folding model analysis.
The situation with the $3^-_1$ state is quite different. The folded inelastic
FF given by the original AMD transition density for the $3^-_1$ state 
overestimates the data in both the DWBA and CC calculations, especially, in  
the standard DWBA1 calculation that uses the same OP for both the entrance and 
exit channels. Using the folded FF rescaled to give a good description of the 
data in the CC calculation, the DWBA1 results still overestimate the data at both 
energies (see Fig.~\ref{inel3}). This is obviously the reason why the best-fit 
$B(E3)$ values given by the (DWBA-based) MDA of the \aap data 
\cite{Itoh11,John03} are much lower than that deduced from the $(e,e')$ data 
(see Table~\ref{t1}). The more accurate DWBA2 and CC calculations, using the 
complex folded OP of the exit channel determined explicitly at the energy 
$E_\alpha-Q$ with the AMD diagonal density of $^{12}$C$^*$, describe the 
\aap data for the $3^-_1$ state much better. 
In this case, the rescaled AMD transition density gives the best-fit 
$B(E3;3^-_1\rightarrow 0^+_1)\approx 60\ e^2$~fm$^6$, which is closer 
to that deduced from the $(e,e')$ data \cite{Kib02}. Note that if the 
inelastic $3^-_1$ form factor is rescaled to fit the data by the DWBA1 
calculation then the rescaled AMD transition density gives the 
best-fit $B(E3;3^-_1\rightarrow 0^+_1)\approx 45\ e^2$~fm$^6$. 
A straightforward explanation is that the $3^-_1$ state is more dilute, 
with the radius given by the diagonal density $\langle r^2 \rangle^{1/2}
\approx 3.14$ fm compared to that of about 2.33 fm for the ground state. 
As a result, the complex folded OP for the exit channel is more absorptive 
at the surface and the strength of the elastic distorted waves is significantly 
reduced. This leads to a reduction of the calculated \aap cross section as 
shown in Fig.~\ref{inel3}. If one rescales, in a similar manner, the 
AMD transition density to give the $B(E3;3^-_1\rightarrow 0^+_1)$ value 
of about 87 $e^2$~fm$^6$ deduced from the $(e,e')$ data, then the absorption 
of the exit channel needs to be further increased in order to describe the 
\aap data for the $3^-_1$ state by either the DWBA2 or CC calculation. 
Such an important effect by the absorption in the exit channel 
has been discussed earlier in more details \cite{Kho08k}. 
 
\begin{figure}
 \begin{center}\vspace*{-1cm}\hspace*{-1cm}
\includegraphics[angle=0,scale=0.8]{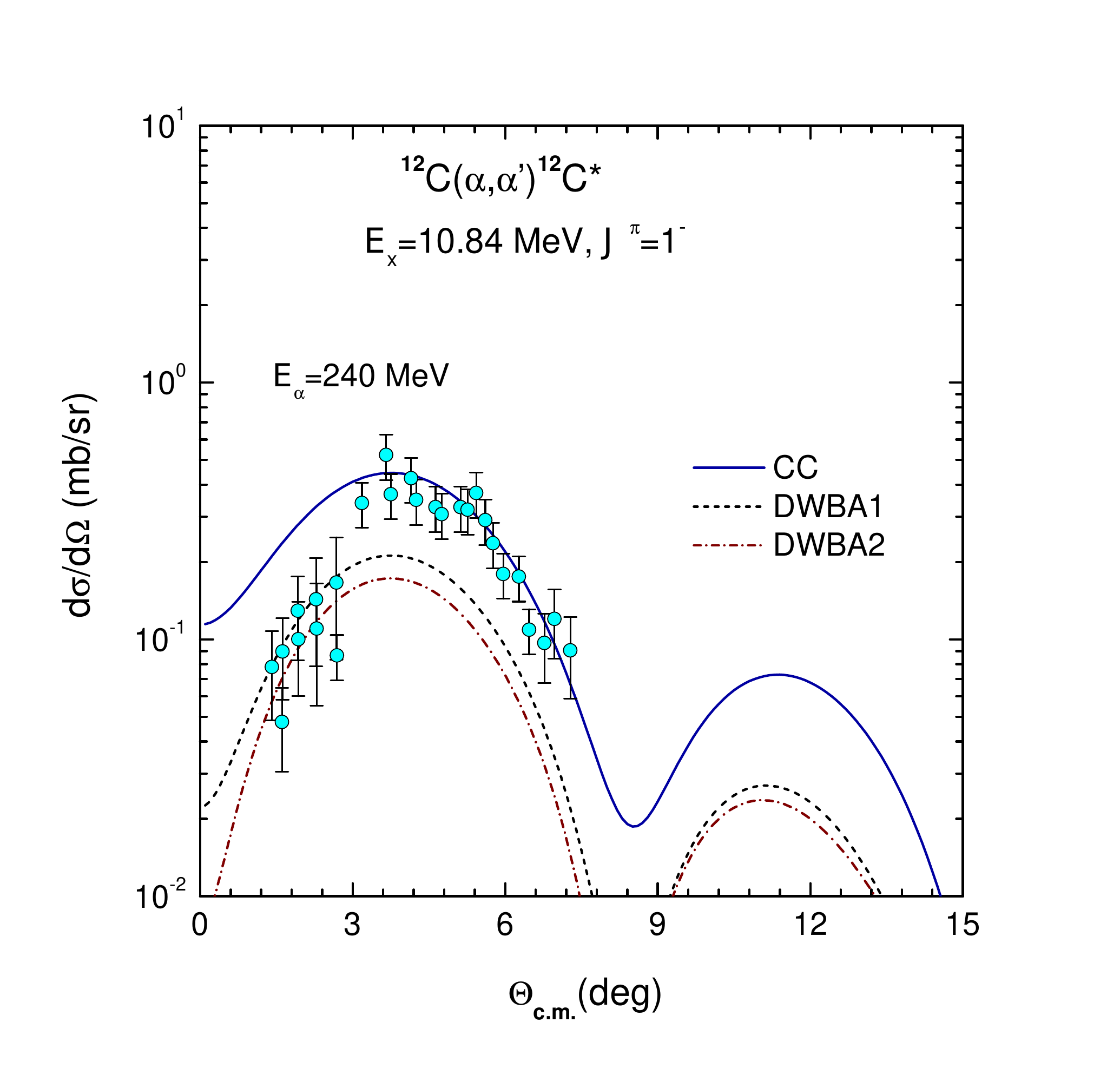}\vspace{-1cm}
\caption{(Color online) The same as Fig.~\ref{inel2} but for the $1^-_1$ state.}
\label{inel1}
 \end{center}
\end{figure}
The isoscalar dipole $1^-_1$ state at $E_x\approx 10.84$ MeV has been 
observed in both \aap experiments at $E_\alpha=240$ and 386 MeV. The 
total \aap angular distribution for the $1^-_1$ state has been deduced by the 
MDA of the 240 MeV data \cite{John03}, covering the first diffraction maximum 
as shown in Fig.~\ref{inel1}. The calculated $1^-$ cross sections have a 
slightly broader bell shape of the first diffraction maximum, and one cannot 
adjust the FF strength to fit all the data points. Like the DWBA analysis 
of Ref.~\cite{John03}, we have tried to obtain the best CC fit of the 
calculated $1^-$ cross section to the data points at the peak of the 
diffraction maximum that have smaller uncertainties.
This procedure implied a renormalization of the AMD transition density that
gives $M(E1;1^-_1\rightarrow 0^+_1)\approx 0.34\ e$~fm$^3$, quite close 
to the DWBA results of Ref.~\cite{John03} given by a collective model 
transition density of the $1^-_1$ state. From the DWBA and CC results 
shown in Fig.~\ref{inel1} one can see quite a strong coupling effect 
caused by the indirect excitation of the $1^-_1$ state via the Hoyle
and $2^+_2$ states, as predicted by the AMD (see Table~\ref{t1} and 
Fig.~\ref{cc}). Note that the direct excitation of the $1^-_1$ state 
has been predicted much stronger, with $M(E1\downarrow)\approx 1.58
\ e$~fm$^3$, and the absorption of the exit channel needs to be 
strongly increased \cite{Kho08k} in order to describe the \aap data 
for the $1^-_1$ state using the original AMD transition density.

\begin{figure}
 \begin{center}\vspace*{-1cm}\hspace*{-1cm}
\includegraphics[angle=0,scale=0.8]{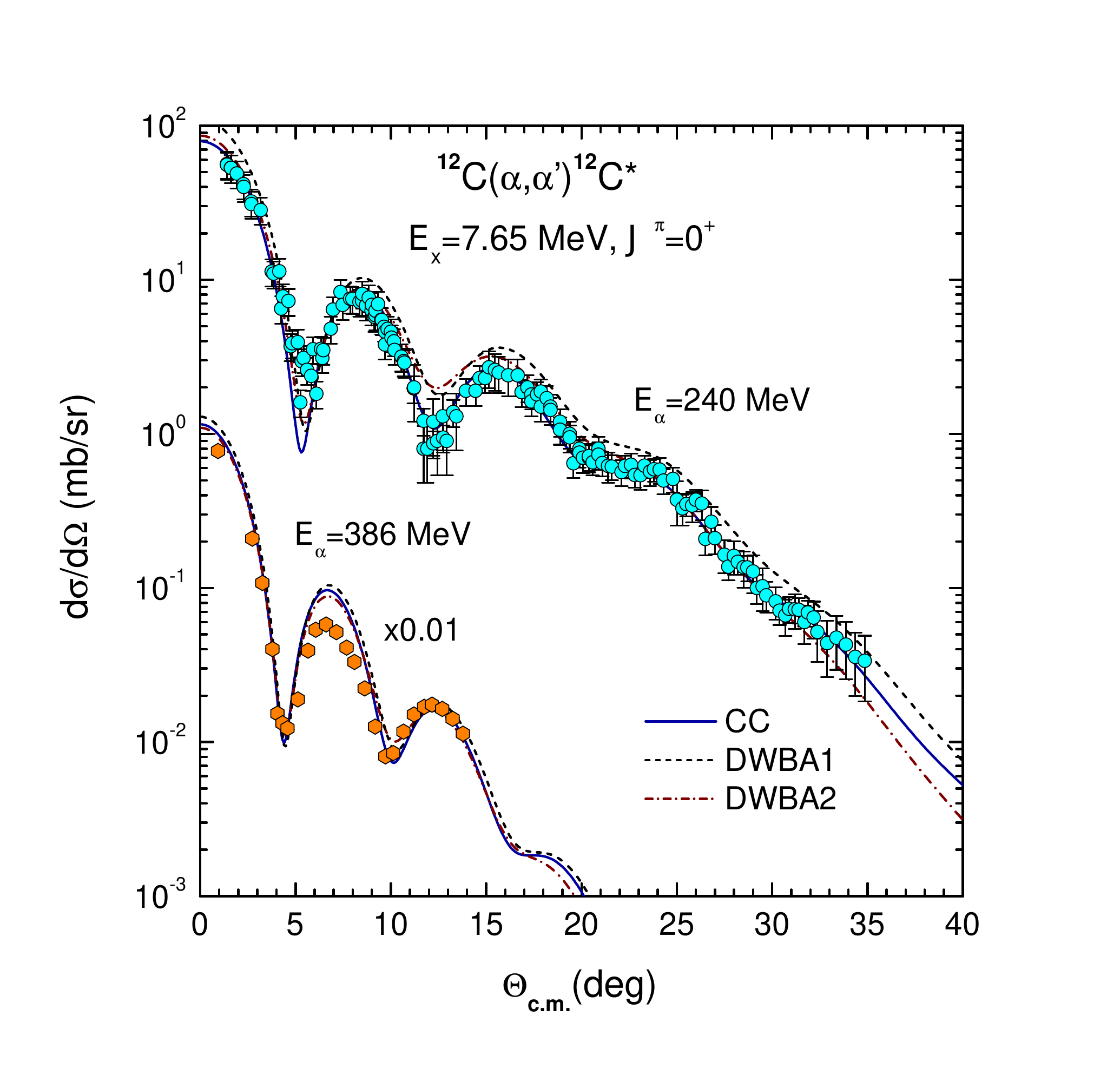}\vspace{-1cm}
\caption{(Color online) The same as Fig.~\ref{inel2} but for the Hoyle ($0^+_2$)
 state.}
\label{inelH}
 \end{center}
\end{figure}
The present folding model + CC analysis of the \aap scattering to the 
Hoyle state (see results shown in Fig.~\ref{inelH}) has revealed interesting 
higher-order coupling effects that are best seen in the results obtained for 
the $\alpha$ energy of 240 MeV. As discussed earlier in Ref.~\cite{Kho08}, 
the MDA of the \aap data measured at different energies has consistently 
found a much weaker $E0$ transition strength of the Hoyle state, with the 
deduced $M(E0;0^+_2\rightarrow 0^+_1)\approx 3.6\sim 3.8\ e$~fm$^2$ that is 
about 30\% weaker than the experimental value $M(E0\downarrow)_{\rm exp}
\approx 5.4\ e$~fm$^2$ deduced from the $(e,e')$ data \cite{Stre70}. 
The DWBA1 calculation using the (rescaled) AMD transition density would 
give the best-fit $M(E0\downarrow)\approx 3.65\ e$~fm$^2$, about the same 
as that given by the RGM transition density rescaled to fit the \aap data 
in the DWBA \cite{Kho08}. The present CC calculation included all possible 
second-order transitions from the Hoyle state to the neighboring cluster states 
(see Fig.~\ref{cc}). In particular, very strong $E\lambda$ transitions between
the Hoyle state and the $3^-_1$, $0^+_3$ and $1^-_1$ states have been taken 
into account (see Table~\ref{t1}). As a result, the best-fit $E0$ strength 
given by the folding model + CC analysis of the 240 MeV data is 
$M(E0\downarrow)\approx 4.5\ e$~fm$^2$ which is about 20\% stronger than 
that given by the standard DWBA analysis. It is likely that a full coupled 
reaction channel analysis of the \aap data including different breakup channels 
would yield the best-fit $M(E0\downarrow)$ value closer to the $(e,e')$ data, 
and that would physically explain the missing monopole strength of the Hoyle 
state in \aap scattering that can be accounted for in the DWBA only by an 
enhanced absorption in the exit channel \cite{Kho08,Kho08k}. 

\begin{figure}
 \begin{center}\vspace*{-1cm}\hspace*{-1cm}
\includegraphics[angle=0,scale=0.7]{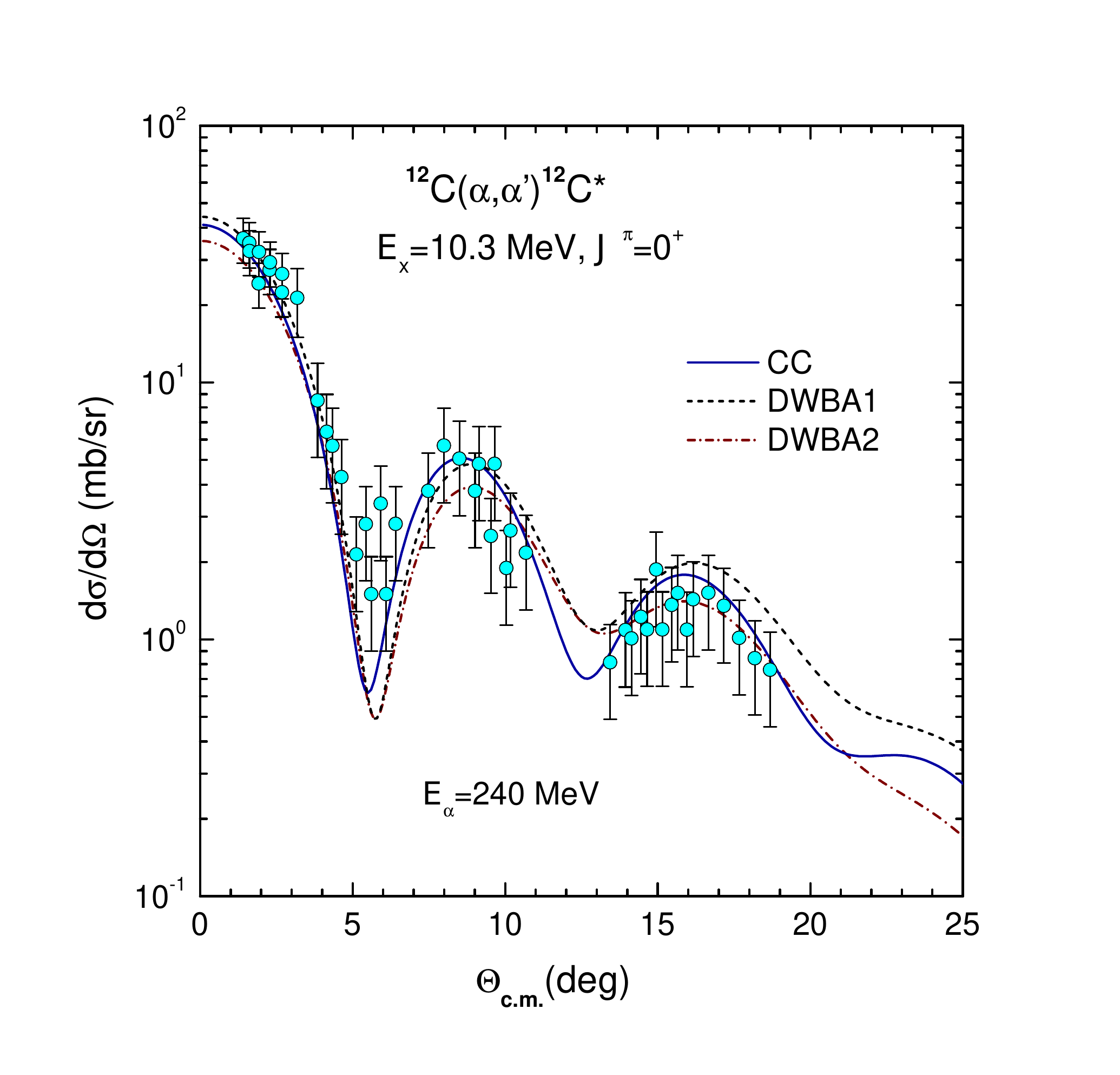}\vspace{-1cm}
\caption{(Color online) DWBA and CC descriptions of the inelastic \aC 
scattering data for the $0^+_3$ state, measured at $E_\alpha= 240$ MeV 
\cite{John03}. The curves DWBA1 and DWBA2 were obtained in the same way as
 described in the caption of Fig.~\ref{inel2}.}\label{inel03}
 \end{center}
\end{figure}
The $E0$ transition strength of the $0^+_3$ resonance has not been unambiguously
determined by previous studies. For example, the RGM calculation predicted about 
the same $M(E0\downarrow)$ value for both the Hoyle and $0^+_3$ states 
\cite{Kam81,Itoh11}, while the AMD gives a much weaker monopole strength of the 
$0^+_3$ state, with the ratio 
$M(E0;0^+_3\rightarrow 0^+_1)/M(E0;0^+_2\rightarrow 0^+_1)\approx 0.34$. 
The multipole decomposition analysis of the 386 MeV data \cite{Itoh11}, using 
the nuclear transition densities from the collective model, give 
$M(E0;0^+_3\rightarrow 0^+_1)/M(E0;0^+_2\rightarrow 0^+_1)\approx 1$. 
Given the CC scheme of Fig.~\ref{cc}, a more precise determination of the $E0$
strength of the $0^+_3$ state should be important for a realistic determination 
of the $E2$ strength of the $2^+_2$ state because these two cluster states are 
connected by a very strong ``interband" $E2$ transition with 
$B(E2;0^+_3\rightarrow 2^+_2)$ predicted by the AMD to be 
around 1500 $e^2$fm$^4$ (see Table~\ref{t1}). Our folding model + DWBA (CC)
analysis of the 240 MeV data for the $0^+_3$ state has been done in the same
manner as discussed above for the $3^-_1$ and Hoyle states, and the results
are plotted in Fig.~\ref{inel03}. The best-fit $E0$ transition strength 
$M(E0;0^+_3\rightarrow 0^+_1)\approx 2.9\ e$~fm$^2$ is close to that given 
by the MDA of the 240 MeV data \cite{John03}. The \aap cross section 
calculated in the CC formalism agrees perfectly with the measured data over 
the whole angular range (see solid curve in Fig.~\ref{inel03}). The DWBA1 
calculation using the same inelastic FF as that used in the CC calculation 
gives a poorer description of the data points at large angles, like the DWBA 
results of Ref.~\cite{John03}. The DWBA2 calculation improves the agreement 
of the calculate $0^+_3$ cross section with the data, but the fit is still 
worse than that given by the CC calculation. Thus, the best description 
of the \aap data measured at $E_\alpha=240$ MeV for both the Hoyle and 
$0^+_3$ states has been consistently given by the present folding model 
+ CC analysis, using the AMD transition densities rescaled to give 
$M(E0\downarrow)\approx 4.5$ and 2.9 $e$~fm$^2$, respectively. 
This result gives the ratio 
$M(E0;0^+_3\rightarrow 0^+_1)/M(E0;0^+_2\rightarrow 0^+_1)\approx 0.64$. 

\begin{figure}
 \begin{center}\vspace*{-1cm}\hspace*{0cm}
\includegraphics[angle=0,scale=0.8]{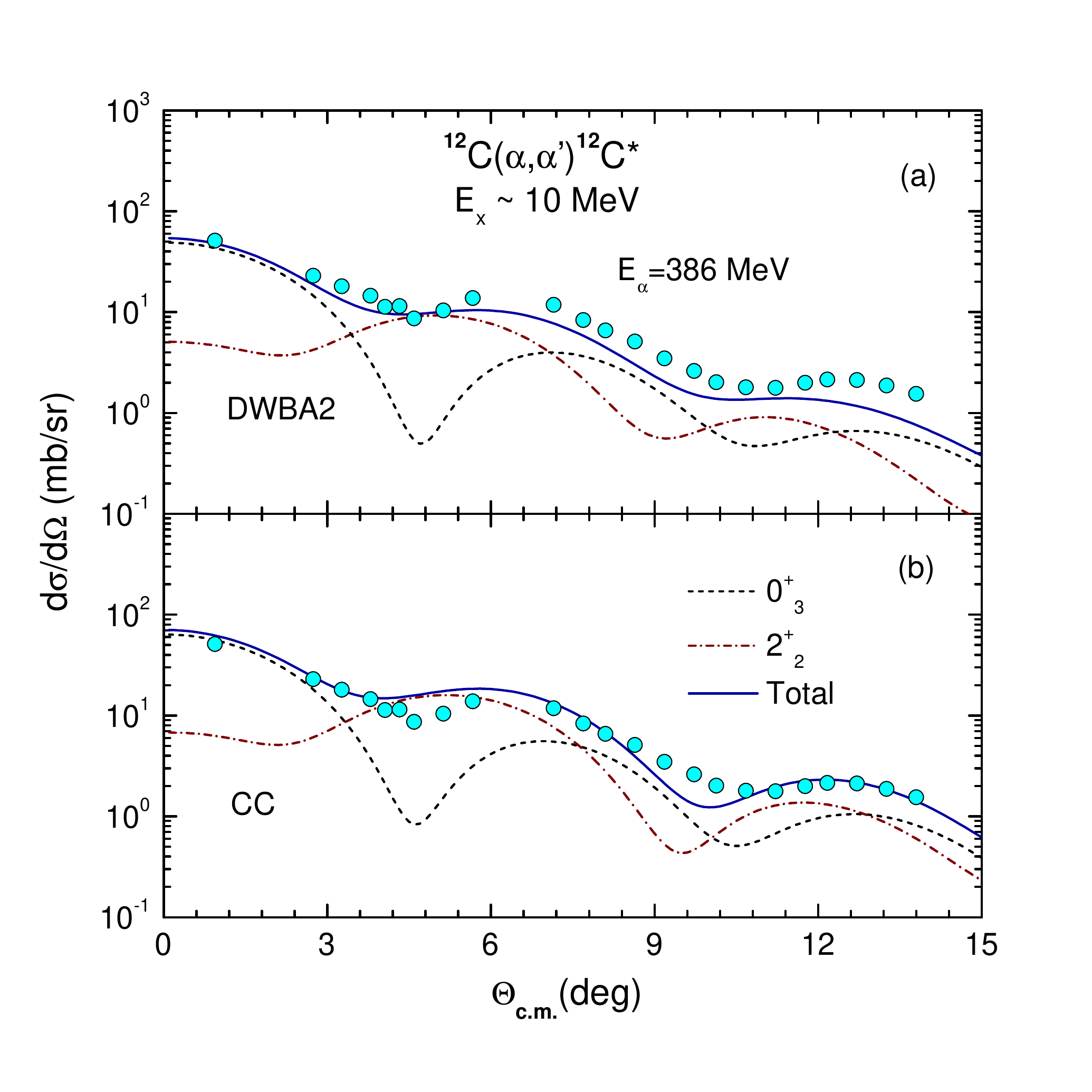}\vspace{-1cm}
\caption{(Color online) DWBA (a) and CC (b) descriptions of the inelastic 
\aC scattering data measured at $E_\alpha= 386$ MeV for the $0^+_3$ and 
$2^+_2$ states \cite{Itoh11}. The DWBA2 and CC results were obtained 
in the same way as described in the caption of Fig.~\ref{inel2}. The
$E2$ strength of the $2^+_2$ state has been adjusted to the best CC fit 
to the data, giving $B(E2\downarrow)\approx 0.6\ e^2$fm$^4$. }
\label{inel386}
 \end{center}
\end{figure}
The MDA of the \aap data measured at $E_\alpha=386$ MeV has shown a broad 
$0^+_3$ resonance and a narrower $2^+_2$ state centered at the excitation 
energies $E_x\approx 9.93$ and 9.84 MeV, respectively. After the subtraction
of the known $0^+_2,\ 3^-_1,$ and $1^-_1$ peaks, the total \aap angular 
distribution deduced for the wide bump centered at $E_x\approx 10$ MeV 
has been shown \cite{Itoh11} to contain only the coherent contributions 
from the $2^+_2$ and $0^+_3$ states (see Fig.~\ref{inel386}). Given the 
$E0$ strength of the $0^+_3$ state accurately determined above in the analysis 
of the 240 MeV data, the $E2$ strength of the $2^+_2$ state remains the only 
parameter in the present analysis of the 386 MeV \aap data. Thus, the strength 
of the $2^+_2$ folded FF was adjusted to the best CC description of the 
\aap angular distribution, as shown in Fig.~\ref{inel386}. Although the 
$\alpha$ energy of 386 MeV can be considered as high enough for the validity 
of the DWBA, very strong $E\lambda$ transitions between the $2^+_2$ state 
and other cluster states of $^{12}$C (see Table~\ref{t1} and Fig.~\ref{cc}) 
have lead to quite a significant coupled channel effect.
From the DWBA2 and CC results shown in Fig.~\ref{inel386} one can see
that the calculated \aap cross section for the $2^+_2$ state is indeed 
enhanced by the indirect excitation of the $2^+_2$ state via other cluster 
states. As a result, the best description of the \aap data measured 
at $E_\alpha=386$ MeV for the $2^+_2$ and $0^+_3$ states is given by the 
folding model + CC calculation using the $2^+_2$ transition density 
rescaled to give $B(E2;2^+_2\rightarrow 0^+_1)\approx 0.6\  e^2$fm$^4$, 
which is about 50\% larger than that predicted by the AMD calculation 
(see Table~\ref{t1}). Although in a fine agreement with  
$B(E2\downarrow)_{\rm exp}\approx 0.73\  e^2$fm$^4$ given by the original
analysis of the photodissociation data \cite{Zim13}, the best-fit 
$B(E2\downarrow)$ value of about 0.6 $e^2$fm$^4$ turns out to be 
significantly lower than $B(E2\downarrow)_{\rm exp}\approx 1.57\  e^2$fm$^4$, 
a value deduced from the revised analysis of the 
$^{12}$C($\gamma,\alpha)^8$Be data \cite{Zim13d}. 
On the other hand, if one sticks to a simple DWBA scenario like that 
in Ref.~\cite{Itoh11} and adjust the $E0$ strength of the $0^+_3$ state 
to fit the data shown in Fig.~\ref{inel386}, keeping the $B(E2\downarrow)$ 
transition rate of the $2^+_2$ state fixed at a value around 
0.4 $e^2$fm$^4$, then the best-fit $E0$ strength of the $0^+_3$ state 
would increase and the agreement between the calculation and experiment 
shown in Fig.~\ref{inel03} would deteriorate. Thus, a consistent folding 
model + CC description of the \aap data measured at both energies 
$E_\alpha=240$ and 386 MeV shown in Figs.~\ref{inel03} and \ref{inel386}, 
respectively, has been achieved with the AMD transition densities rescaled 
to give the best-fit $E2$ and $E0$ strengths tabulated in Table~\ref{t1} 
for the $2^+_2$ and $0^+_3$ states.     

\begin{figure}
 \begin{center}\vspace*{-1.5cm}\hspace*{0cm}
\includegraphics[angle=0,scale=0.8]{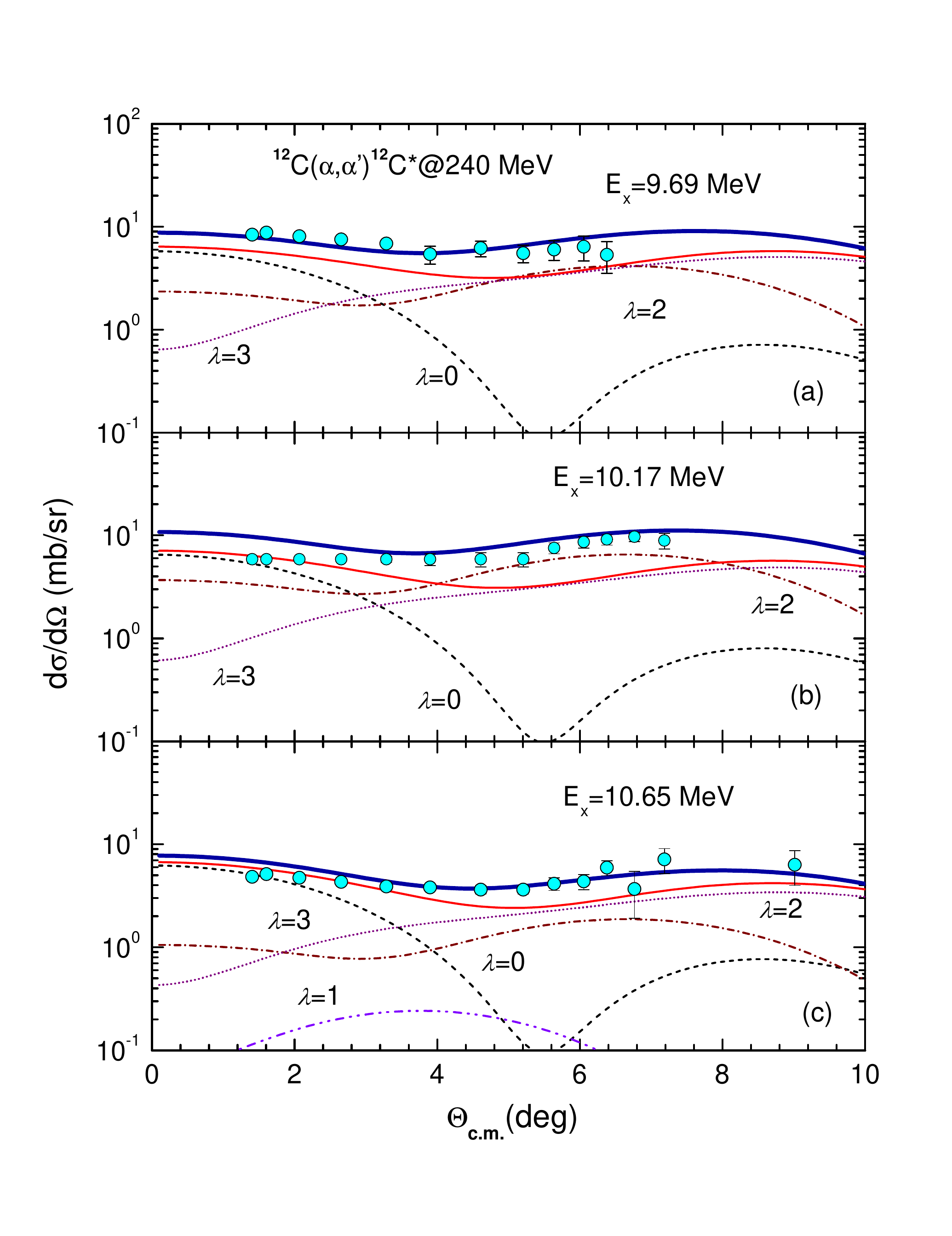}\vspace{-1cm}
\caption{(Color online) Differential \aap cross sections measured at 
$E_\alpha=240$ MeV  \cite{John03,John11} for the 475 keV-wide energy bins 
centered at $E_x=9.69$ MeV (a), 10.17 MeV (b), and 10.65 MeV (c), and the 
CC results given by the contributions of different $2^\lambda$-pole 
transition strengths. The total cross sections obtained with and without 
the contribution from the $2^+_2$ state are shown as the thick (blue) and 
thin (red) solid lines, respectively.} \label{intotal240}
 \end{center}
\end{figure}
A natural question now is why the $2^+_2$ state has not been observed at 
$E_x\approx 10$ MeV in the \aap experiment at $E_\alpha=240$ MeV. In fact, 
the MDA of the 240 MeV \aap data has established a $2^+$ peak at $E_x\approx 
11.46$ MeV, with the width of about 430 keV and $B(E2\downarrow)\approx 0.5
\ e^2$fm$^4$, that could be assigned to the $2^+_2$ state \cite{John03}. 
Given the realistic $E\lambda$ strengths of the isoscalar states found
above in our folding model + CC analysis of both data sets, we decided to look 
again at the 240 MeV \aap data measured for several energy bins around 10 MeV
\cite{John03,John11}. As discussed in Sec.~\ref{amd}, the $3^-$ transition 
strength found in each energy bin by the MDA of the 240 MeV \aap data 
\cite{John03} was used to scale the AMD transition density to obtain the 
$3^-$ inelastic FF of the bin. The best-fit $E\lambda$ transition strengths 
found above for the $2^+_2,\ 0^+_3$ and $1^-_1$ states were distributed over 
the energy bins by the averaging procedure (\ref{av1})-(\ref{av3}) for the 
determination of the corresponding inelastic FF of the bin. We note
that a width $\Gamma=315$ keV \cite{Fre09} has been assumed in the averaging 
of the $E1$ strength of the $1^-_1$ state. The CC description of the 
240 MeV \aap data measured for three energy bins closest to $E_x=10$ MeV is 
shown in Fig.~\ref{intotal240}. From the calculated total cross section with 
(thick solid lines) and without the contribution from the $2^+_2$ state 
(thin solid lines) one can see clearly that the $E2$ strength of the $2^+_2$ 
state is indeed present in these energy bins. Because the CC description of 
the \aap data shown in Fig.~\ref{intotal240} has been obtained without any 
further readjusting the $E\lambda$ strengths of the involved cluster states, 
we conclude that the presence of the $2^+_2$ state at the energy near 10 MeV 
has been found by the present folding model + CC analysis of the \aap data 
at $E_\alpha=240$ MeV. Such a subtle effect could not be resolved in the 
original MDA of the 240 MeV \aap data.          

\begin{figure}
 \begin{center}\vspace*{-1.5cm}\hspace*{0cm}
\includegraphics[angle=0,scale=0.8]{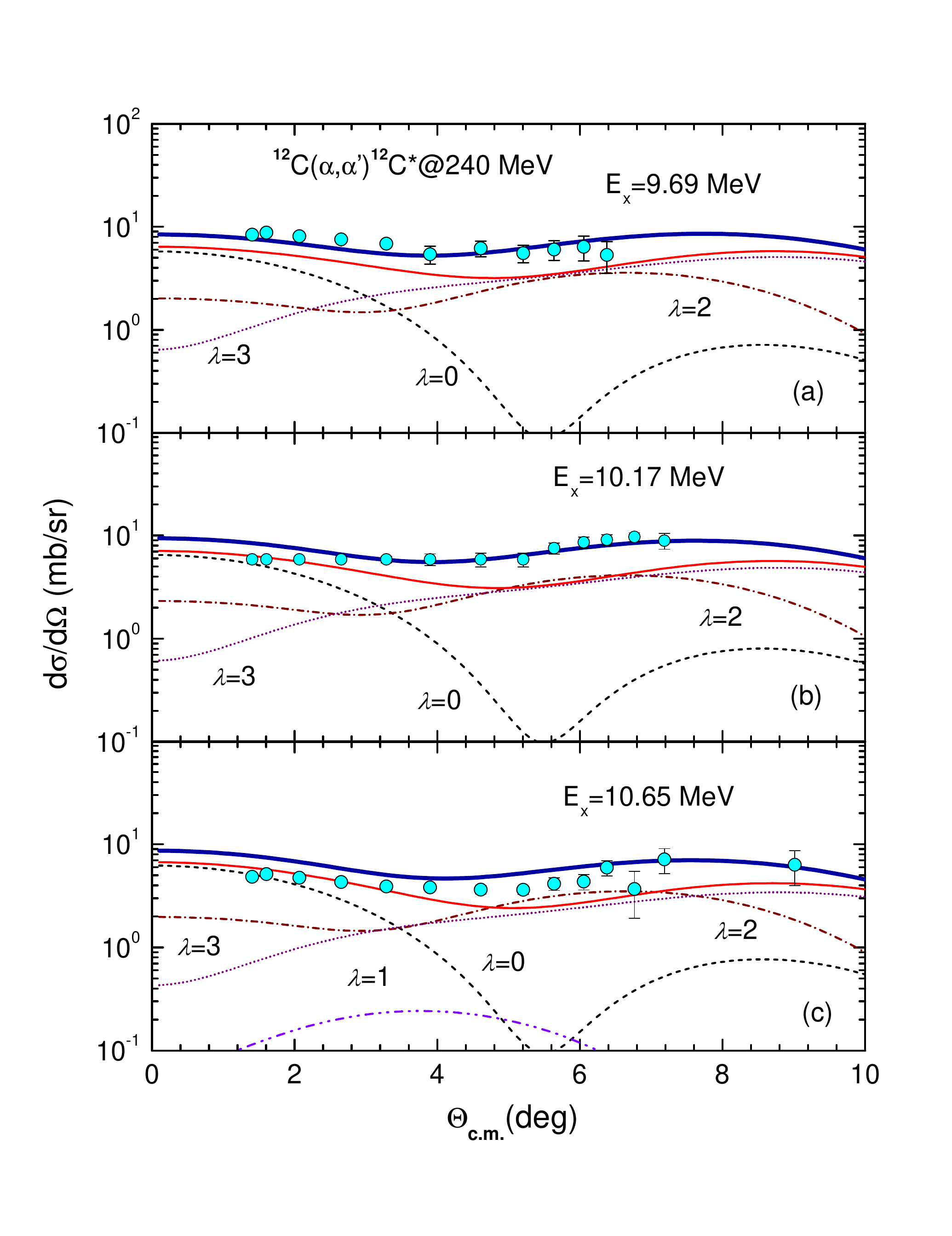}\vspace{-1cm}
\caption{(Color online) The same as Fig.~\ref{intotal240} but with the
$E2$ strength of the $2^+_2$ state distributed over a wider energy range
spanned by the total width $\Gamma=2.1$ MeV.} \label{intotal240b}
 \end{center}
\end{figure}
Due to the uncertainty in the measured total width of the $2^+_2$ state
discussed above in Sec.~\ref{amd}, the $E2$ strength distribution of this
state was built up in two distinct scenarios by the averaging procedure
(\ref{av1}) using the total width $\Gamma=0.8$ and 2.1 MeV for this state. 
To assess the effect by the width of the $E2$ strength distribution, we 
have calculated again the 240 MeV \aap cross sections for the same three 
energy bins as in Fig.~\ref{intotal240} but using the $E2$ strength of the 
$2^+_2$ state distributed over a wider energy range spanned by the total 
width $\Gamma=2.1$ MeV. From the results shown in Fig.~\ref{intotal240b}  
one can see a better agreement of the CC results with the data in the 
energy bins centered at $E_x=9.69$ and 10.17 MeV, while the agreement 
with the data in the bin centered at 10.65 MeV slightly worsens. 
A clear presence of the $2^+_2$ state at the energy near 10 MeV can 
also be seen in Fig.~\ref{intotal240b}, which consistently confirms our 
conclusion on the $2^+_2$ peak in the 240 MeV \aap spectrum.        

\begin{figure}
 \begin{center}\vspace*{-1.5cm}\hspace*{0cm}
\includegraphics[angle=0,scale=0.8]{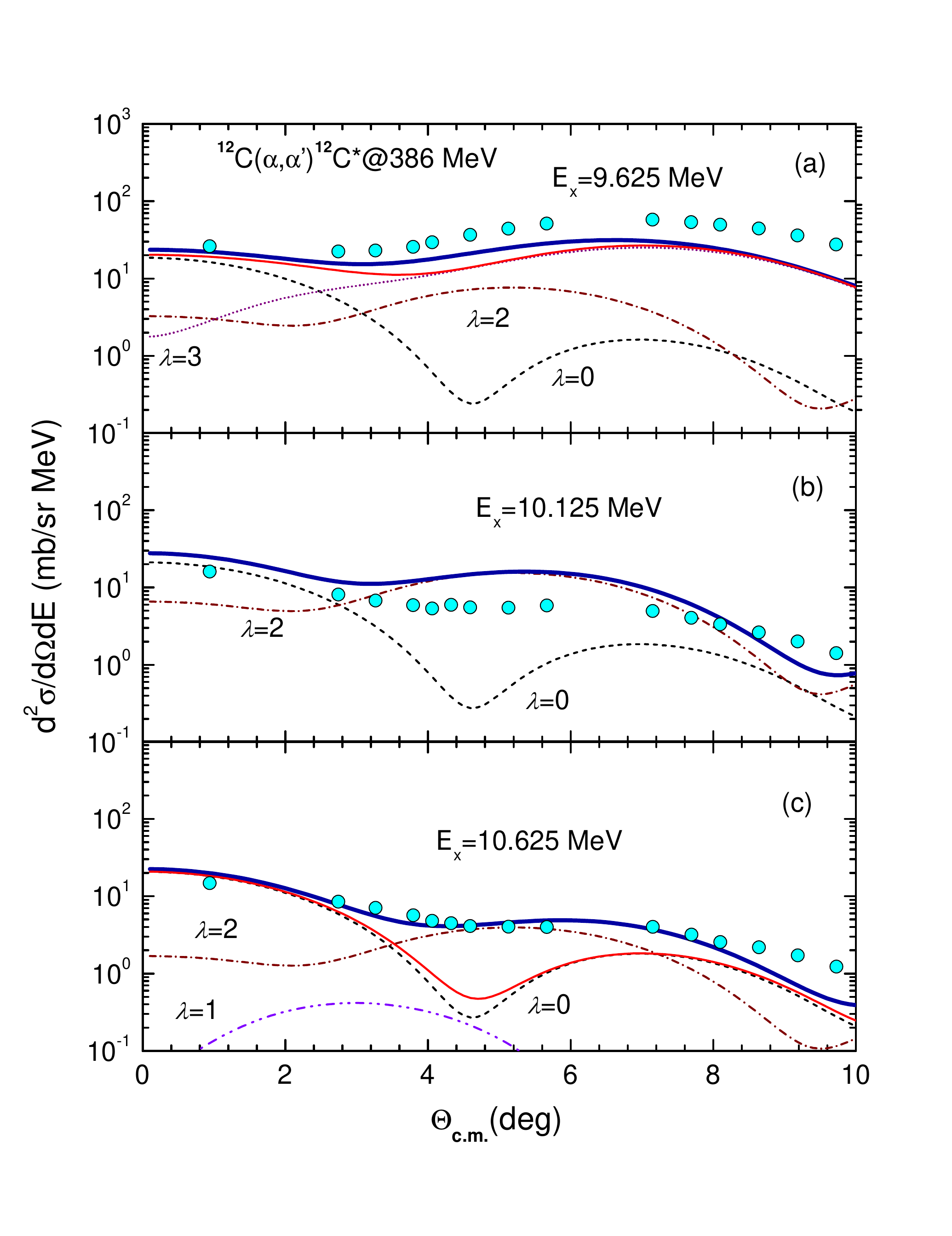}\vspace{-1cm}
\caption{(Color online) Double-differential \aap cross sections measured at 
$E_\alpha=386$ MeV  \cite{Itoh11} for the 250 keV-wide energy bins centered at 
$E_x=9.625$ MeV (a), 10.125 MeV (b), and 10.625 MeV (c), in comparison with 
the CC results in the same way as in Fig.~\ref{intotal240}.}
\label{intotal386a}
 \end{center}
\end{figure}
\begin{figure}
 \begin{center}\vspace*{-1.5cm}\hspace*{0cm}
\includegraphics[angle=0,scale=0.8]{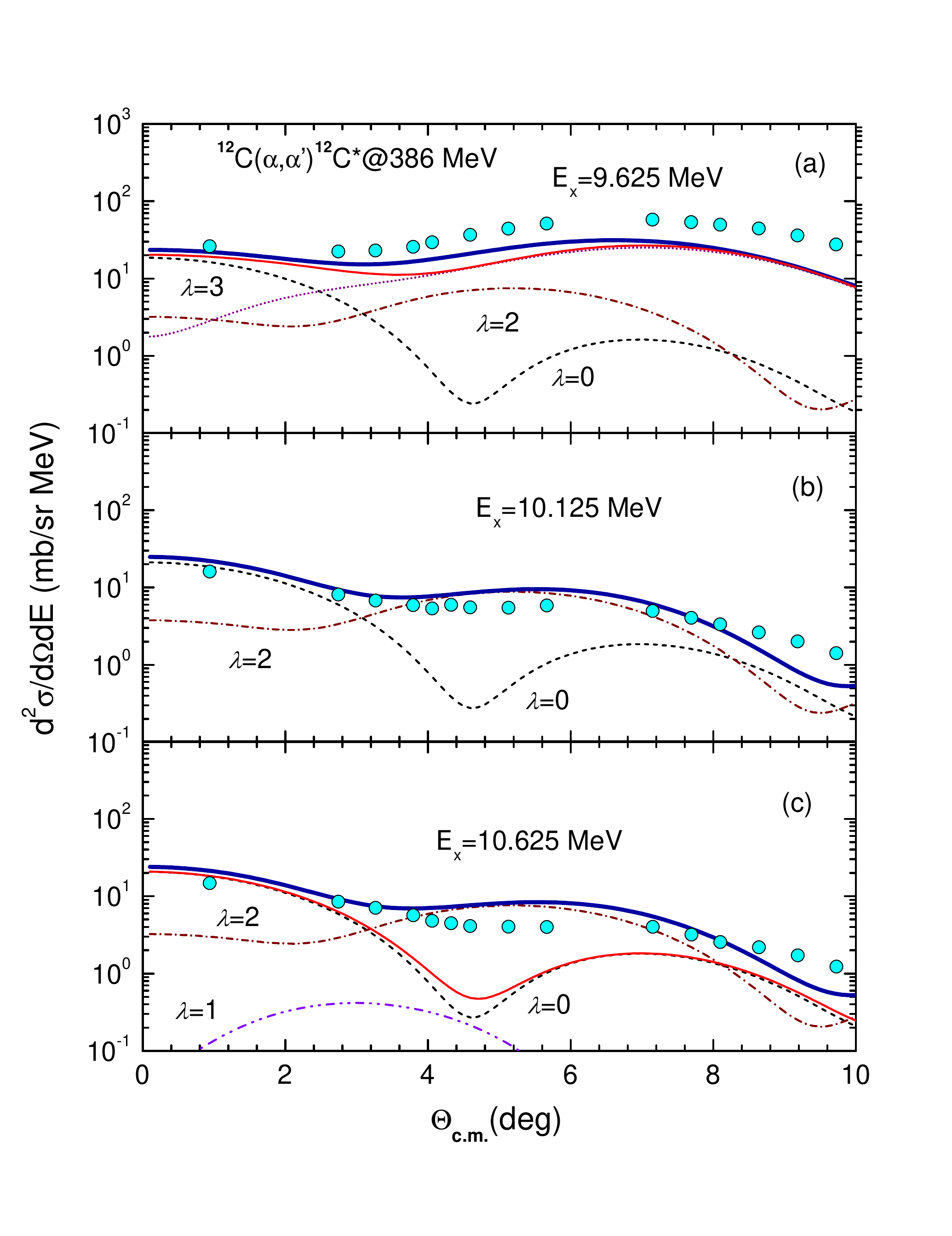}\vspace{-1cm}
\caption{(Color online) The same as Fig.~\ref{intotal386a} but with the
$E2$ strength of the $2^+_2$ state distributed over a wider energy range
spanned by the total width $\Gamma=2.1$ MeV.}
\label{intotal386b}
 \end{center}
\end{figure}

The inelastic \aC scattering at $E_\alpha=386$ MeV was measured using
the high-precision Grand Raiden spectrometer, and the \aap spectrum 
over the whole energy and angular range has been obtained free of background 
\cite{Itoh11}. In difference from the MDA of the 240 MeV data, the MDA of the 
386 MeV data has revealed a clear presence of the $2^+_2$ state at the energy 
near 10 MeV, and the total \aap cross section measured at this energy was used 
above in our analysis to determine the realistic $E2$ strength of the $2^+_2$ 
state (see Fig.~\ref{inel386} and the discussion thereafter). 
With the 386 MeV data available also for many energy bins around $E_x=10$ MeV, 
it is of interest to probe the consistency of the present folding model + CC 
approach in the calculation of \aap scattering at $E_\alpha=386$ MeV, similar 
to that shown in Figs.~\ref{intotal240} and \ref{intotal240b}. We note that 
the $3^-$ transition strength found in the energy bins around 10 MeV by the 
MDA of the 386 MeV data \cite{Itoh11} is better resolved in energy than that 
found by the MDA of the 240 MeV data \cite{John03}, and it was used to scale 
the AMD transition density to obtain the folded $3^-$ inelastic FF of the bin. 
All the remaining inputs of the folding model + CC calculation were determined 
in the same manner as that done above for the 240 MeV data. The CC description 
of the 386 MeV \aap data measured for three similar energy bins around 
$E_x=10$ MeV is shown in Fig.~\ref{intotal386a}. One can see that a good 
overall agreement of our results with the \aap data measured 
at $E_\alpha=386$ MeV for these energy bins has been achieved using the same 
structure inputs for the most important cluster states of $^{12}$C as those
used to obtain the CC results shown in Fig.~\ref{intotal240}. The CC 
results for the same three energy bins obtained with the $E2$ strength 
of the $2^+_2$ state distributed over a wider energy range spanned by the total 
width $\Gamma=2.1$ MeV are shown in Fig.~\ref{intotal386b}, and one can also
see a good agreement of the CC results with the data, especially, a better CC
description of the data taken for the energy bin centered at $E_x=10.125$ MeV. 
This result might well indicate that the $2^+_2$ state has indeed a wide total 
width $\Gamma\approx 2$ MeV. 

\section{Summary}

The complex OP and inelastic scattering FF given by the double-folding model
using the nuclear densities predicted by the AMD approach and the CDM3Y6 
interaction have been used in the comprehensive CC calculation of the elastic 
and inelastic \aC scattering at $E_\alpha=240$ and 386 MeV. The \aap cross 
sections calculated in the CC approach for the (isoscalar) cluster states 
of $^{12}$C are compared with the \aap data under study, and the strength 
of the inelastic FF has been fine tuned in each case to the best CC description 
of the measured angular distribution, to determine the corresponding 
$E\lambda$ transition strength. A detailed folding model + CC analysis 
of the \aap data measured in the energy bins around $E_x\approx 10$ MeV 
has been carried out to reveal the $E2$ transition strength that can be 
assigned to the $2^+_2$ state of $^{12}$C.   

A clear presence of the $2^+_2$ state of $^{12}$C at the excitation energy  
$E_x\approx 10$ MeV has been confirmed consistently by the present analysis 
of the \aap data measured at $E_\alpha=240$ and 386 MeV. 
Given quite strong $E\lambda$ strengths predicted for the $E\lambda$ 
transitions between the $2^+_2$ state and other cluster states 
of $^{12}$C, a high-precision \aap measurement at the lower beam energy 
might be an interesting alternative to observe the $2^+_2$ excitation 
and to probe the indirect (two-step) excitation of this state via the 
CC scheme shown in Fig.~\ref{cc}.  

The obtained best-fit $E\lambda$ strengths of the considered states agree 
reasonably with the existing database, with the exception of the 
$B(E2)_{\rm exp}$ transition rate of the $2^+_2$ state given by the 
revised analysis of the $^{12}$C($\gamma,\alpha)^8$Be data \cite{Zim13d} 
that is more than double the best-fit $B(E2)$ value found in our analysis. 
This result stresses the need for new precise measurements of the excitation
of $^{12}$C using the $\alpha$ beam as well as other probes.    
Some difference between the $E\lambda$ transition strengths of the 
$0^+_2,\ 0^+_3,\ 3^-_1$, and $2^+_2$ states given by the present analysis 
and those given by the earlier multipole decomposition analyses of the 
same \aap data \cite{Itoh11,John03} has been shown to be due, in part, 
to the strong coupled channel effect and enhanced absorption in the 
exit channel of the \aap scattering. 

\section*{Acknowledgements}
We thank M.~Itoh, B.~John, T.~Kawabata, and W.R.~Zimmerman for their helpful
communications. The present research has been supported by the National 
Foundation for Science and Technology Development (NAFOSTED) 
under Project No. 103.04-2011.21.

\end{document}